\documentclass[dvips,12pt,a4paper]{article}
\usepackage{epsfig}
\usepackage{a4wide}
\usepackage{amssymb}
\begin{document}
\pagestyle{empty}
\vspace{1. truecm}
\begin{center}
\begin{Large}
{\bf Multiplicity Distribution studies of

 $e^+e^-$-annihilation at 50-61.4 and at
 172-189 GeV by Two Stage Model hadronization
 \footnote
{Talk given at the X1 Annual Seminar
"Nonlinear Phenomena in Complex Systems"
Minsk, Belarus, May 13-16, 2002}}

\end{Large}

\vspace{2.0cm}
{\large  E.S. Kokoulina}
\\[0.3cm]
The P.O. Sukhoi State Technical University of Gomel, 246746  Belarus
\end{center}
\begin{center}
\begin{small}
{$e-mail:helen_ {-}by@yahoo.com$}
\end{small}
\end{center}
\vspace{.5cm}

\begin{abstract}
\noindent
The multiplicity distributions at high energy $e^+e^-$~
annihilation are described  well within the Two Stage Model
in the region from $14$ to $189$ GeV. Energy dependence
of  parameters of this model gives the dynamic
picture of the parton stage and the stage of
hadronization.
It is shown that oscillations in sign of the ratio
of factorial cumulant moments over factorial moments
of increasing order can be confirmed by this model.

\vspace*{3.0mm}

\noindent
\end{abstract}


\pagestyle{plain}

\section{Introduction}

Multiparticle production (MP) is one of the most
important topics in high energy physics. Using MP
we can get more information about the nature of strong
interactions and understand deeper the structure of
matter. Over the last few years many thorough reviews devoted to
MP have been done \cite{SIS}. Modern accelerators
have  made it possible to study MP more extensively
and in detail. Developing theory of high energy physics
quantum chromodynamics~(QCD) \cite{POL}
and a lot of phenomenologic models are tested by the
process of MP.

Multiparticle processes begin at high energy. Among all
producing particles we can observe a lot of hadrons.
On one hand we want to know high energy
physics, but on the other hand the increase of the
inelastic channels makes it difficult to
describe this process with customary methods.
The situation concerning the history of thermodynamics
developing and statistical physics is much the same.
Analysis of MP process is carried out using of statistical
methods because the number of secondaries in $e^+e^-$ annihilation
is large (more than 60) \cite{OPA}. The consideration of MP
begins from the behavior of charged multiplicity.

As it is generally known the multiplicity is the number of
secondaries  {\bf n} in process of MP:
${\bf A}  + {\bf B} \rightarrow {\bf a_1}+{\bf a_2}+
\ldots +{\bf a_n}$.
The multiplicity distribution (MD) $\bf P_n$
is the ratio of cross-sections  ${\sigma_n}$  to
$\sigma=\sum\limits_n\sigma_n$:
$
\bf{P_n}=\sigma_n/\sigma.
$
This quantity has the following meaning: the probability
of producing of $\bf n$ charged particles in
this process. We can also construct
quantities such as mean values, moments of
MD, can study correlations and so on.

Investigation of MP has led to discovery of
jets. Jets phenomena can be studied in
all processes, where  energetic partons are
produced. The most common ones are in $e^+e^-$
annihilation, deep inelastic scattering
of $e$, $\mu$ or $\nu$ on nucleons and hadron-
hadron scattering, involving high-$\bf p_T$
particles in final state.
Let us consider  $e^+e^-$ -annihilation
at high energy. This process is one of the
most suitable for the study of MP.
In accordance with QCD it can be
realized through the production of
$\gamma$ or $Z^0$--boson into two quarks:

\begin{equation}
\label{1}
e^+e^-\rightarrow(Z^0/\gamma)\rightarrow q\bar q
\end{equation}

Perturbative QCD can describe the process of fission partons
(quarks and gluons) at high energy, because
the strong coupling $\alpha_s$ is
small at that energy. This stage can be called as
the stage of cascade. After hard fission, when
partons have not high energy, they must be
changed into hadrons, which we can observe.
On this stage we shouldn't apply perturbative
QCD. Therefore phenomenological models are used for
description of hadronization (transformation of
quarks and gluons into hadrons) in this case.

The description of the stage quark-gluon
cascade by means of perturbative QCD was applied
in \cite{{KUV},{DDT}}. Certain features
of the predictions at the parton level
are expected to be insensitive to
details of the hadronization mechanism.
They were tested directly by using
hadron distributions \cite{HOY}.

The $e^+e^-$--reaction is simple for analysis,
as the produced state is pure $q\overline q$.
It is usually difficult to determine the
quark species on event-by-event basis. The
experimental results are averaged over the
quark type. Because of confinement the
produced quark and gluons fragment into
jets of observable hadrons.

The hadronization models are more
phenomenological and are built on the
experience gained from the study of low--$\bf{p_T}$
hadron collisions.It is usually considered that the
producing of hadrons from partons is
universal process.

\section{Two Stage Model}

Parton spectra in QCD quark and gluon jets
were studied by Konishi~K., Ukawa~A. and
Veneciano~G.\cite{KUV}. Working at the leading
logarithm approximation and avoiding IR
divergences by considering finite $x$, the
probabilistic nature of the problem has
been established \cite{KUV}.

At the studying of MP at high energy we
used idea of A.~Giovannini~\cite{GIO} for
description of quark-gluon jets as Markov branching
processes. Giovannini proposed to interpret
the natural QCD evolution parameter $\bf {Y}$
\begin{equation}
\label{2}
{\bf {Y}}=\frac{1}{2\pi b}\ln[1+\alpha_{s} b\ln(\frac{Q^2}{\mu^2})] ,
\end{equation}
where $2\pi b=\frac{1}{6}(11N_C-2N_f)$ for a theory
with $N_C$ colours and $N_f$ flavours,
as the thickness of the jets and their
development as Markov process.

Three elementary processes contribute
into QCD jets:\\
(1) gluon fission;\\
(2) quark bremsstrahlung;\\
(3) quark pair creation.

Let $A\Delta Y$ be the probability that
gluon in the infinitesimal interval $\Delta Y$
will convert into two gluons, $\widetilde{A}\Delta Y$
be the probability that  quark will radiate a gluon,
and $B\Delta Y$ be the probability that
a quark-antiquark pair will be created from
a gluon. $A, \widetilde A, B$ are assumed to be
{\bf Y}-independent constants and each individual
parton acts independently from others, always
with the same infinitesimal probability.

Let us define the probability that parton will
be transformed into $m$~gluons over a jet of
Y in thickness and call it $P_{m}^{P}(Y)$.
The probability generating function for
a parton jet will be

\begin{equation}
\label{3}
Q^{P}(z;Y)=\sum\limits_{m=0}^{\infty}
P_{m}^{P}(Y)z^{m}.
\end{equation}

A.Giovannini constructed system of
differential equations and obtained
explicit solutions of MD for a parton
jet in particular case $B=0$ (process
of quark pair creation is absent)
In the common case $B\neq0$ MD are
similar to particular one \cite{GIO}.

For quark jet explicit solutions are
given \cite{GIO}
$$P_{0}(Y)=e^{-\widetilde AY},$$

\begin{equation}
\label{4}
P_{m}(Y)=\frac{\mu(\mu+1)\dots
(\mu+m-1)}{m!}e^{-\widetilde AY}
(1-e^{-AY})^{m},
\end{equation}
where $\mu=\frac{\widetilde A}{A}$ .
Futher the average gluon multiplicity is
$\overline m=\mu (e^{AY}-1)$ and the
normalized exclusive cross section
for producing $m$ gluons from quark is
$$
\frac{\sigma_{m}^{q}}{\sigma_{tot}}\equiv
P_{m}(Y)=
$$
\begin{equation}
\label{5}
=\frac{\mu(\mu+1)\dots(\mu+m-1)}{m!}
\left[\frac{\overline m}{\overline m+\mu}\right]^{m}
\left[\frac{\mu}{\overline m+\mu}\right]^{\mu}.
\end{equation}

The generating function (\ref{3}) will be given by
\begin{equation}
\label{6}
Q^{q}(z,{\bf Y})=\sum\limits_{m=0}^{\infty}
z^{m} P_{m}(Y)=
 \left[\frac{e^{-AY}}{1-z(1-e^{-AY})}
\right]^{\mu}.
\end{equation}
Eq.(\ref{4}) is Polya-Egenberger
distribution, where $\mu$ is non-integer.

In Two Stage Model \cite{TSM} we took
(\ref{4}) for description of cascade
stage and added supernarrow binomial
distribution for hadronization stage.
We chose it based ourselves on analysis
of experimental data in $e^+e^-$-
annihilation lower 9 GeV. Second
correlation moments were negative
at this energy. The choise of such
distributions was the only
could describe experiment.

We  suppose  that hypothesis
of soft colourless is right.
We add stage of hadronization to
parton stage with aid of it's
factorization. MD in this
process can be written
\begin{equation}
\label{7}
P_n(s)=\sum\limits_mP^P_mP_n^H(m,s),
\end{equation}
where $P_m^P$ is MD for partons
(\ref{4}), $P_n^H(m,s)$
- MD for hadrons produced from
$m$ partons on the stage of
hadronization. Futher we will use
instead of parameter $\bf Y$ CM(center of
masses) energy
$\sqrt s$.

In accordance with TSM the stage
of hard fission of partons
is described by negative binomial
distribution (NBD) for quark jet
\begin{equation}
\label{8}
P_m^P(s)=\frac{k_p(k_p+1)\dots(k_p+
m-1)}{m!}\left(\frac{\overline m}
{\overline m+k_p}\right)^{m}\left(
\frac{k_p}{k_p+\overline m}\right)^{k_p},
\end{equation}
where $k_p=\widetilde A/A$, {\quad}
$\overline m=\sum\limits_mmP_m^P$.
We neglect process (3) quark
pair production ($B=0$). Two quarks
fracture to partons independently
of one another. Total MD of two quarks
is equal to (\ref{7}) too. Parameters
$k_p$ and ${\overline m}$ of MD for two
joint quark-antiguark jets are
doubled, but we use that designations.

 $P_m^P$ and generating function
for MD $Q^P(s,z)$ are
\begin{equation}
\label{9}
P_m^P=\frac{1}{m!}\frac{\partial^m}
{\partial z^m}\left.Q^P(s,z)\right|_{z=0},
\end{equation}
\begin{equation}
\label{10}
Q_m^P(s,z)=\left[1+\frac{\overline m}
{k_p}(1-z)\right]^{-k_p}.
\end{equation}
MD of hadrons formed from parton are described
in form \cite{TSM}
\begin{equation}
\label{11}
P_n^H=C^n_{N_p}\left(\frac{\overline n^h_p}
{N_p}\right)^n\left(1-\frac{\overline n_p^h}
{N_p}\right)^{N_p-n},
\end{equation}
($C_{N_p}^n$ - binomial coefficient)
with generating function
\begin{equation}
\label{12}
Q^H_p=\left[1+\frac{\overline n^h_p}
{N_p}(z-1)\right]^{N_p},
\end{equation}
where $\overline n^h_p$ and $N_p$
($p=q,g$) have meaning of average
multiplicity and maximum secondaries
of hadrons are formed from parton
on the stage of hadronization.
MD of hadrons in $e^+e^-$
annihilation are determined by
convolution of two stages
\begin{equation}
\label{13}
P_n(s)=\sum\limits_{m=o}^{\infty}
P_m^P\frac{\partial^n}{\partial z^n}
(Q^H)^{2+m}|_{z=0},
\end{equation}
where $2+m$ is total number of partons
(two quarks and $m$ gluons) .

Further we do the following simplification
for the second stage: $\frac{\overline n^h_q}
{N_q}\approx\frac{\overline n^h_g}
{N_g}$, considering that probabilities
of formation of hadron from quark or gluon
are equal. We introduce parameter
$\alpha=\frac{N_g}{N_q}$ for distinguishing
between hadron jets, created from quark or
gluon on the second stage. We also make
simplification for designation
$N=N_q$, $\overline n^h=\overline n^h_q$.
Then we get
$$
Q^H_q=\left(1+\frac{\overline n^h}
{N}(z-1)\right)^N,
$$
$$Q^H_g=\left(1+\frac{\overline n^h}
{N}(z-1)\right)^{\alpha N}.
$$
Introducing in (\ref{13}) expressions
(\ref{8}), (\ref{12}) and differentiating
on $z$ we obtain MD of hadrons in the process
of $e^+e^-$ annihilation in TSM
\begin{equation}
\label{14}
P_n(s)=\sum\limits_{m=0}P_m^P
C^n_{2+\alpha m)N}\left(\frac{\overline n^h}
{N}\right)^n\left(1-\frac{\overline n^h}
{N}\right)^{(2+\alpha m)N-n}.
\end{equation}
For comparing with experimental
data the normalized factor $\Omega$ was
introduced into (\ref{13}) and a number
of gluons in the sum was restricted
by $M_g$ - maximal number of possible
gluons created on the first stage
\begin{equation}
\label{15}
P_n(s)=\Omega \sum\limits_{m=0}
^{M_g}P_m^PC^n_{(2+\alpha m)N}
\left(\frac{\overline n^h}
{N}\right)^n\left(1-\frac
{\overline n^h}{N}\right)^{(2
+\alpha m)N-n}.
\end{equation}

The results of comparison of
model expression (\ref{15})
with experimental data \cite{DAT} are
represented in Table 1 (parameters
of two stages) and on Figures (1)-(18).
We can see that MD in TSM
(solid curve) are describing well
the experimental $e^+e^-$-data
(black square $\blacksquare$) from $14$
to $189$ GeV. Summing up is
limited to $M_g$ equal $20-22$ for energies
to $61.4$GeV and $33-41$ above,
because further increase does not
change $\chi ^2$.

The description of experimental
MD \cite{DAT} by (\ref {15}) gives
$\chi ^2$ $\sim $ 1 for the most
energies with the expection of
$34.8$GeV and $55$GeV (more high $\chi ^2$),
$N_{DF}=14$.
Calculated MD are giving small
deviations from experimental data
at $34.8$GeV, but for almost
all multiplicities (Fig.3).
Probably it is connected with
small statistics of events.

At $55$GeV they give big
deviations, but only to the right side
of central region (Fig.7).
It can be described as energy suppression of
formation of hard gluons connected
with appearance of heavy $q\bar{q}$ pair
or heavy hadrons.
Energy behavior of total cross section of
$e^+e^-\rightarrow$ hadrons \cite{TOT}
is changed to a sharp rise from here.
It also may be connected with the reason of
narrowing of distribution.

\newpage
\begin{center}
Table 1. Parameters of TSM.
\end{center}
\renewcommand{\tablename}{Table}
\begin{center}
\begin{tabular}{||c||c|c|c|c|c|c||c||}
\hline
\hline
$\sqrt s$ GeV &$\overline m$ &$k_p$&$N$&$\overline n^h$&$\alpha $&$\Omega $&$\chi^2$\\
\hline
\hline
14  &  .08 &$2.4\times 10^8$& 27.7 & 2.87 & .97 &2.&2.75\\
\hline
22  & 3.01 & 4.91      & 20.2 & 4.34 & .2  &2.  &1.29\\
\hline
34.8& 6.58 & 6.96      & 12.5 & 4.1  & .195&2.&240 \\
\hline
43.6& 10.3 & 48.3      &  5.16& 2.31 & .444&2.  &5.36\\
\hline
50  &  7.48&  1.3      & 24.6 & 6.14 & .1  &2.09&1.97\\
\hline
52  &  11.5&  1.       & 24.8 & 6.16 & .104&2.46&2.51\\
\hline
55  &8.6&$6.\times 10^4$&17.&   4.   & .26 &2.2 &124\\
\hline
56  &9.81&8.23&           6.51&3.73  & .273&2.02&1.29\\
\hline
57  &11.3&14.4&           4.&  2.76  & .385&2.&1.95\\
\hline
60  &8.92&9.&           9.31&4.2&      .254&1.99&2.97\\
\hline
60.8&9.52&6.68&         7.&  4.12&     .246&2.02&2.98\\
\hline
61.4&10.4& 1.        &21.1&  6.38&     .108&2.29 &1.76\\
\hline
91.4& 10.9 &  7.86     & 11.2 & 4.8  & .226&2.   & 1.16\\
\hline
172 & 20.1 &  9.11     &  9.17& 4.34 & .195&1.98&6.86\\
\hline
183 & 13.2 &  1.48     & 54.6 & 8.9  & .086&2.06&2.56\\
\hline
189 & 15.1 &  6.9      & 11.6 & 5.15 & .215&2.01&2.37\\
\hline
\hline
\end{tabular}
\end{center}

\section{Dynamics of multiparticle production}
We will analyse dynamics of MP which
corresponds to values of
parameters of TSM $\qquad$ (Table~1).
We will begin from a cascade stage.
This stage is described by two parameters:
$\overline{m}$ and $k_p$.

The average multiplicity of gluons
$\overline m$ formed on fission stage
has tendency to rise. It is
changed from $\sim 0.1$ at
$14$GeV to $\sim 20$ at $183$ Gev.
But we can see certain insignificant
deviation from this direction at
$\sqrt s$=$50-61.4$GeV, $\quad$
at $183$ GeV.
 It followes from QCD that in
particular case ($B=0$) the
parameter $k_p$ that equal
to ratio $ 2\tilde{A}/A \rightarrow 1$.
Values $k_p$ are changed insignificantly.
They are remained $\sim 10$ at
almost all energies.
There is some physical senses
of this parameter.
One of the most interesting from
them is temperature $T$ \cite{TEM}:
$T=k_p^{-1}$. From thermodynamical
models we can also obtain the
following connection

\begin{equation}
\label{16}
k_p^{-1}=T_0+1/cE,
\end{equation}
where $T_0$ is the temperature of
system before interaction, $c$ -
thermal capacity, $E$ - energy
spended on creationg new particles
\cite{TEM}. In this sense we can
make assumptions: temperature of parton
system with developed cascad are
lowest at $14$ and $55GeV$ than
at the others.
The highest temperatures are
reached on first stage at $50$,
$52$, $61.4$ and $183GeV$.

The interesting picture of
hadronization is discovered
in conformity with parameters of
second stage $N_q$, $\overline n_q^h$
and $\alpha$ . The parameter
$N_q$ determines maximum number of
hadrons, which can be formed
from quark on this stage.
In TSM (Table 1) it takes
different values from $4$ to
$55$. We can't reveal steady
energy rise or fall for it.
Big $N_q$ point to predominance of
hadrons formed from quark jets,
small $N_q$ point to essential
contribution gluon jets in the
hadron multiplicity. More
probably that this parameter
remaines constant and $\sim {16}$
with small deviations.

Next parameter $\overline n_q^h$
has meaning of mean hadron multiplicity
from quark on second stage.
We can see the tendency to weak rise
with big scatter.  Such
behavior of parameter may be
connected with the growth
of spectrum of mass hadron states
(appearance of new mass states
with increase of energy).
The average value of
$\overline n_q^h$ is
about $5-6$ in the research
region.

Parameter $\alpha$ was introduced
for comparision quark and gluon
jets. It is almost constant and
equal to $0.2$ with some deviations.
If we know $\alpha$ then
we can determine analogous parameters
$N_g=\alpha N_q$ and $\overline
n_g^h=
\alpha \overline n_q^h$
for gluon jet. It is interesting
that these parameters remain
constant without considerable
deviations: $N_g\sim3$ and
$\overline n_g^h\sim1$ (Figures
19 and 20). From this result
we can affirm about universality
of hadronization.

The fact that $\alpha <1$
says that hadronization of
gluon jets are more soft than
quark one. The simplest explanation
to this phenomenon is the fact
that a quark takes away more
conciderable energy than gluon.

The ratio
$\frac{\overline n^h_q}{N_q}$
determines the probability of
formation of hadron from parton.
It is increased from $\sim .1$
to $\sim$.5 with rise of energy
from $14$ to $43.6$GeV, then we have
big variations in region $50-61.4$
GeV ($.23-.69$) and the ratio is
almost constant at higher energies
($\sim .4$) (Figure 21).
It should be noted that
the small probabilities
are realized at $55$GeV (in the
region $50-61.4$GeV) and at
$183$GeV.

The normalized factor $\Omega$
remains constant and is equal to 2.

\section{Oscillation of moments in MD}
It was shown recently \cite{OSC}
that the ratio of
factorial cumulative moments over
factorial moments  changes sign as
a function of order. We can use
MD formed in TSM for explanation of
this phenomenon.

The factorial moments can be obtained
from MD $P_n$ through the relation
\begin{equation}
\label{17}
F_q=\sum\limits_{n=q}^{\infty}
n(n-1)\dots(n-q+1)P_n ,
\end{equation}
and factorial cumulative moments
are found from expression
\begin{equation}
\label{18}
K_q=F_q-\sum\limits_{i=1}^{q-1}C_{q-i}^{i}
K_{q-i}F_i .
\end{equation}
The ratio of their quantities is
\begin{equation}
\label{19}
H_q=K_q/F_q.
\end{equation}
We can use the generating function for
MD of hadrons (\ref{14}) in $e^+e^-$
annihilation G(z)
$$
G(z)=\sum\limits_{m=0}P_m^g[Q_g^H(z)]^m Q^2_q(z)=
$$
\begin{equation}
\label{20}
=Q^g (Q^H_g(z))^m Q^2_q(z) .
\end{equation}
We are calculating $F_q$ and $K_q$
in TSM using (\ref{20})
\begin{equation}
\label{21}
F_q=\frac{1}{\overline n^q(s)}\left.\frac
{\partial^qG}{\partial z^q}\right|_{z=0}
\end{equation}
\begin{equation}
\label{22}
K_q=\frac{1}{\overline n^q(s)}\left.\frac
{\partial ^q\ln G}{\partial z^q}\right|_{z=0}.
\end{equation}
The expression(\ref{20}) for G(z)
after taking a logarithm
$$
\ln G(s,z)=-k_p\ln [1+\frac{\overline m}{k_p}(1-Q^H_g)]+2\ln Q^H_q
$$
and the expansion to series in power
on $Q^H_g$ will be
\begin{equation}
\label{23}
\ln G(s,z)=k_p\sum\limits_{m=1}\left(
\frac{\overline m}{\overline m+k_p}\right)^m
\frac{Q^m_g}{m}+2\ln Q^H_q.
\end{equation}
Inserting $Q_g$ into (\ref{23})
$$\ln G(s,z)=k_p\sum\limits_{m=0}\left(
\frac{\overline m}{\overline m+k_p}
\right)^m\frac{1}{m}\left[1+\frac{\overline n^h}{N}
(z-1)\right]^{\alpha mN}+
$$
$$
+2N\ln [1+\frac{\overline n^h}{N}(z-1)],
$$
and using (\ref{22}) we obtain
$$
K_q=\left(k_p\sum_{m=1}\alpha m (\alpha m -\frac{1}{N})\dots
(\alpha m-\frac{q-1}{N})\left(\frac{\overline m}
{\overline m+k_p}\right)^m\frac{1}{m}\right.
$$
\begin{equation}
\label{24}
\left.-2(-1)^q\frac{(q-1)!}{N^{q-1}}\right)\left(
\frac{\overline n^h}{\overline n(s)}\right)^q
\end{equation}
where $\overline n(s)$ is the average
multiplicity hadrons in process (\ref{1}).
It is possible to find $F_q$ using (\ref{21})
\begin{equation}
\label{25}
F_q=\sum\limits_{m=0}(2+\alpha m)(2+\alpha m-
\frac{1}{N})\dots(2+\alpha m-\frac{q-1}{N})
P_m\left(\frac{\overline n^h}{\overline n(s)}
\right)^q
\end{equation}
with $P_m$ equal (\ref{8}).

The sought-for expression for $H_q$ will be
\begin{equation}
\label{26}
H_q=\Omega_1\frac{\sum\limits_{m=1}k_p\alpha m
(\alpha m-\frac{1}{N})\dots(\alpha m-\frac{q-1}
{N})(\frac{\overline m}{\overline m+k_p})^m\frac{1}{m}-
2(-1)^q\frac{(q-1)!}{N^{q-1}}}
{\sum\limits_{m=0}(2+\alpha m)(2+\alpha m-
\frac{1}{N})\dots(2+\alpha m-\frac{q-1}{N})P_m}
\end{equation}
where $\Omega_1$ is the normalized factor.
The comparison with experimental data \cite{OSC}
shows that (\ref{26}) describes the ratio of
factorial moments (Figures 22-39). It is seen
minimum at q=5.

In the region before $Z^0$ $H_q$ may
oscillate in sign only with the period equal
$2$, changed sign with parity $q$.
At more high energies the period
is increased to $4$ and higher. It can be
explained by influence of hadronization.
Values $K_q$ (as well $H_q$) may change
sign only owing to second summand in (\ref{24}).
More devoloped cascad of partons with hadronization
comes to big period of oscillations of sign.

The immediate calculations $H_q$ based on
(\ref{17})-(\ref{19}) with using MD(\ref{15})
gives very good description of the
oscillation value of $H_q$ ($\chi^2 \approx2$).
Significant oscillations start near region
producing of $Z^0$ and can be explained by
non-integer values of parameters of hadronization
$N_q$ and $N_g=\alpha N_q$ or by convolution
of wide (for parton jets) and narrow
(for hadron jets on second stage) MD.

\section{Conclusions}
It is shown that TSM does not contradict to
the experimental data on MD and the oscillations
ratio of factorial moments. TSM offeres
concreted physical picture of multiplicity
production in high energy $e^+e^-$ annihilation.

\section*{Acknowledgments}
I am deeply indebted to Kuvshinov V.I. with
whom I worked and developed TSM, to head of
International Center for Advanced Studies at
the P.O. Sukhoi State Technical University of Gomel
Solovtsov I.L., chief of Laboratory of Physical
Investigations Pankov A.A. and to all my
collegues in Gomel State Technical
University for support in my job.

\newpage
\begin{figure}
\begin{minipage}[b]{.3\linewidth}
\centering
\includegraphics[width=\linewidth, height=2in, angle=0]{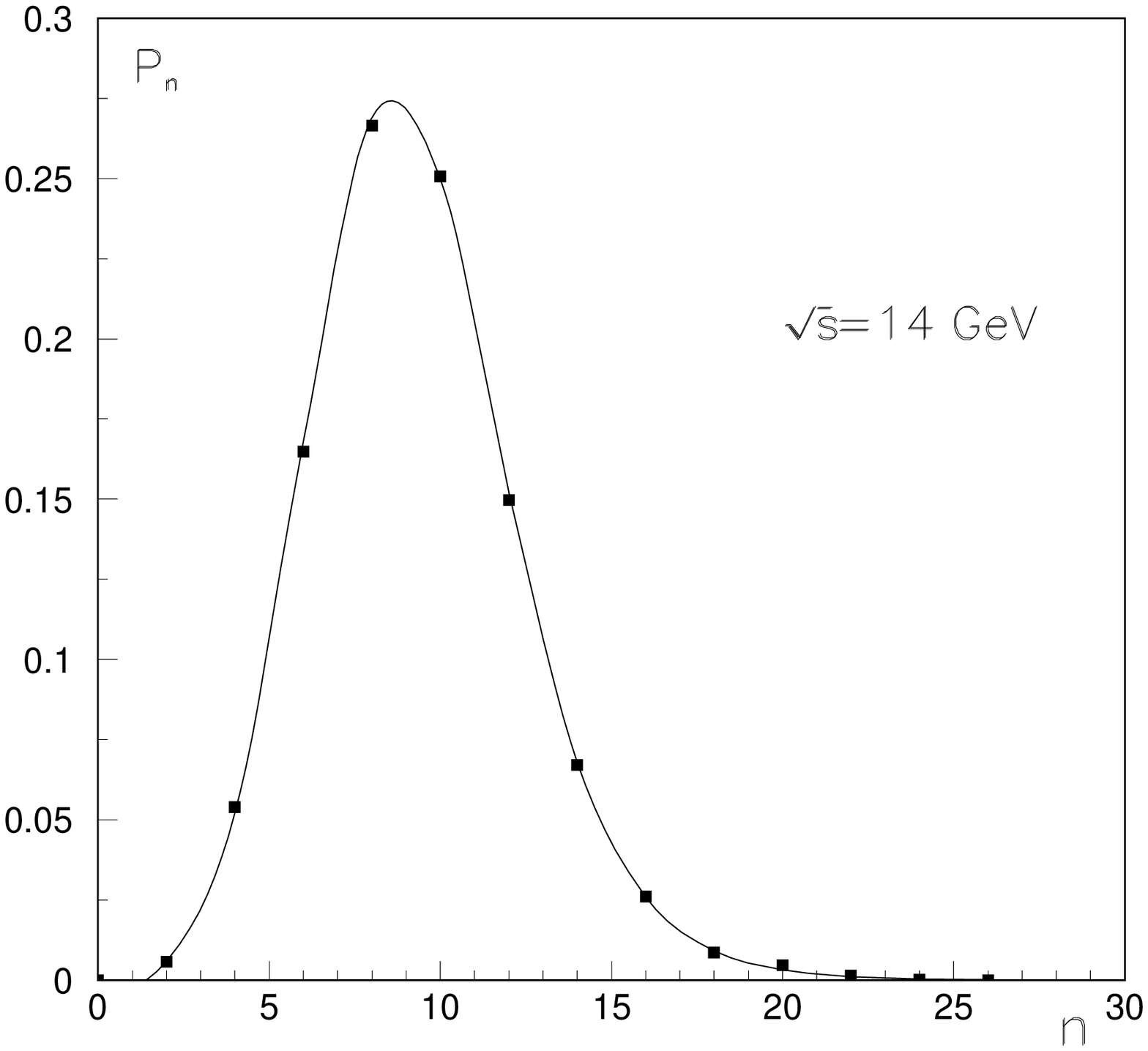}
\caption{MD at 14GeV.}
\label{1dfig}
\end{minipage}\hfill
\begin{minipage}[b]{.3\linewidth}
\centering
\includegraphics[width=\linewidth, height=2in, angle=0]{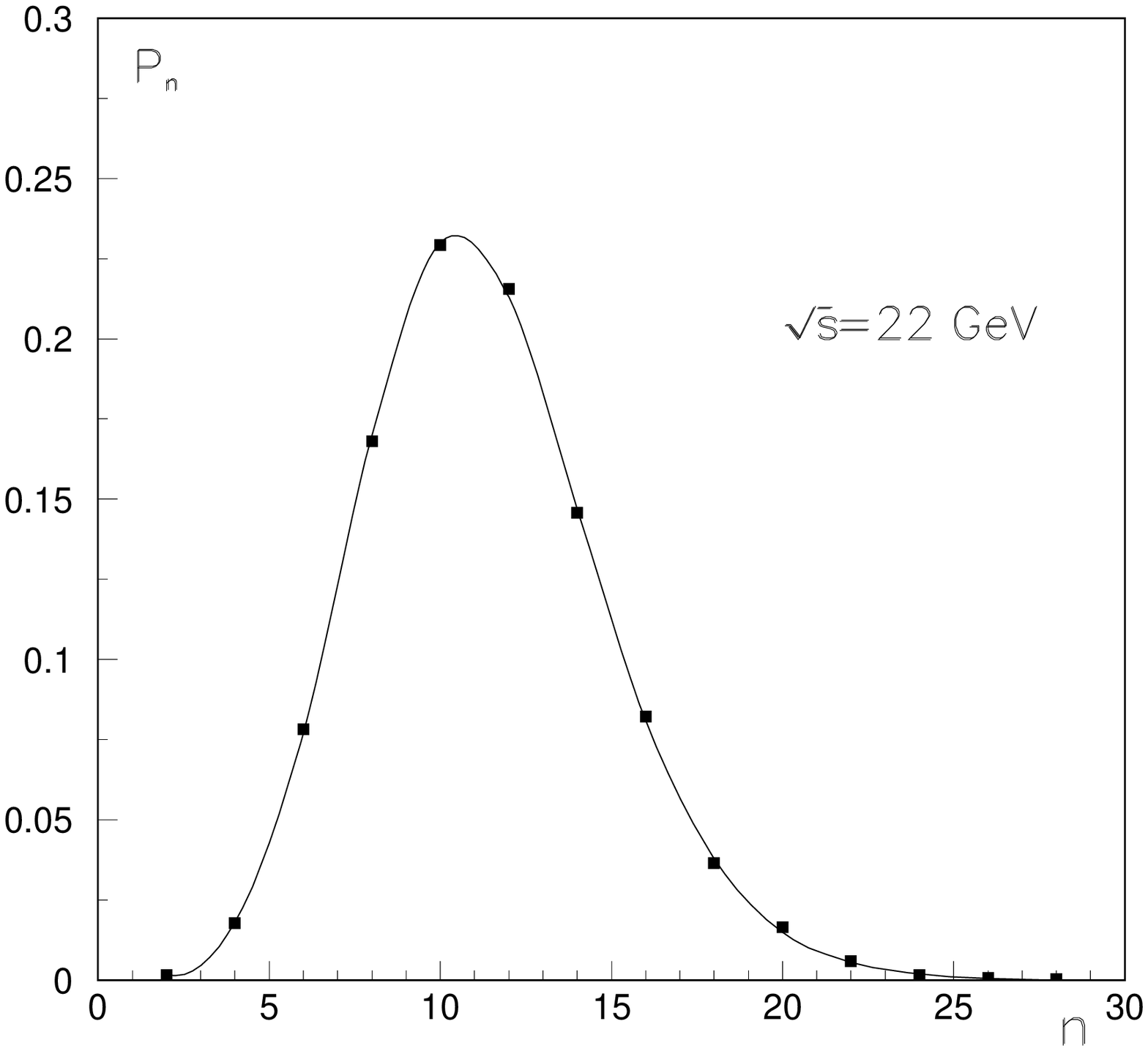}
\caption{MD at 22GeV.}
\label{2dfig}
\end{minipage}\hfill
\begin{minipage}[b]{.3\linewidth}
\centering
\includegraphics[width=\linewidth, height=2in, angle=0]{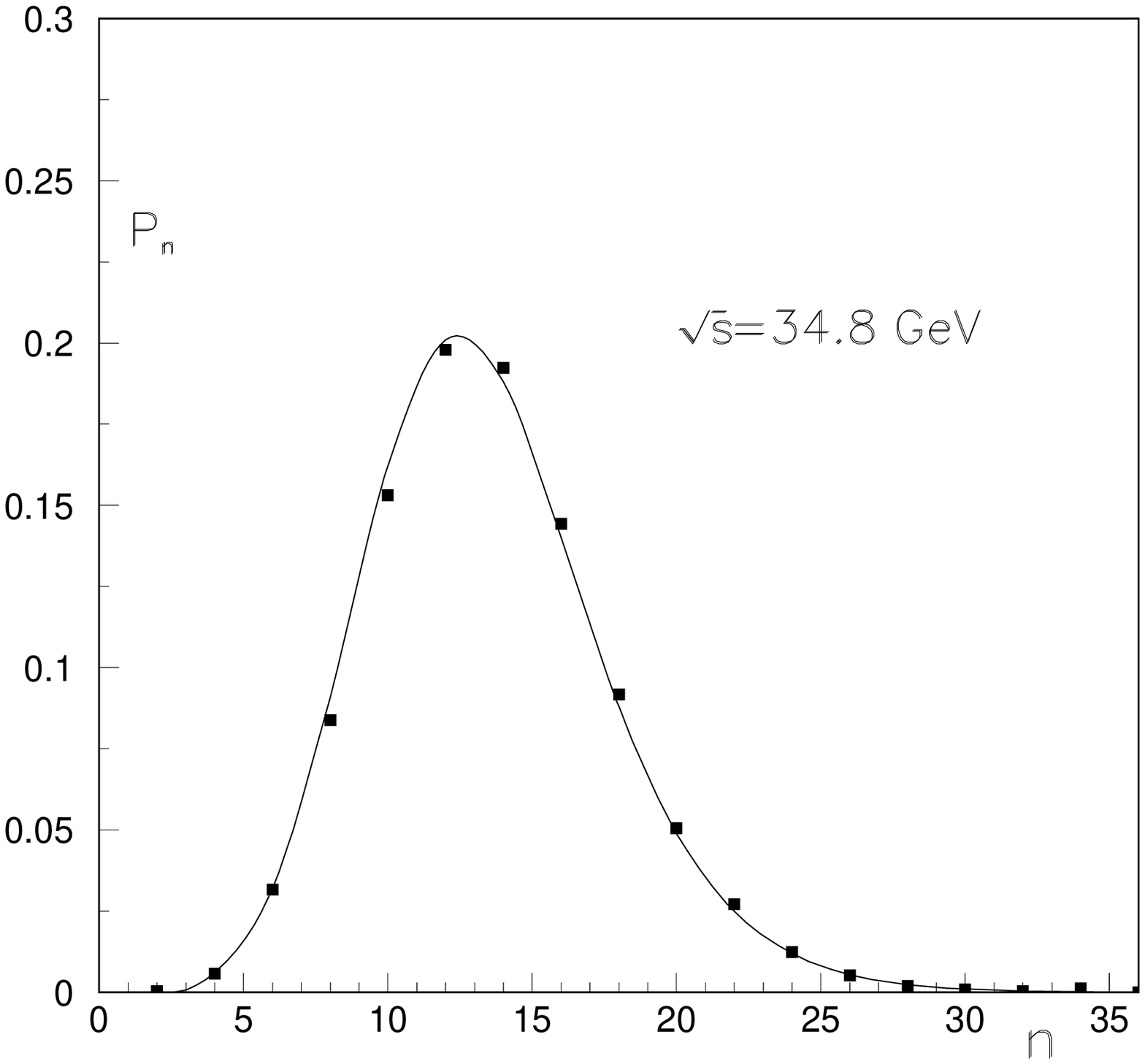}
\caption{MD at 34.8GeV}
\label{3dfig}
\end{minipage}
\end{figure}

\begin{figure}
\begin{minipage}[b]{.3\linewidth}
\centering
\includegraphics[width=\linewidth, height=2in, angle=0]{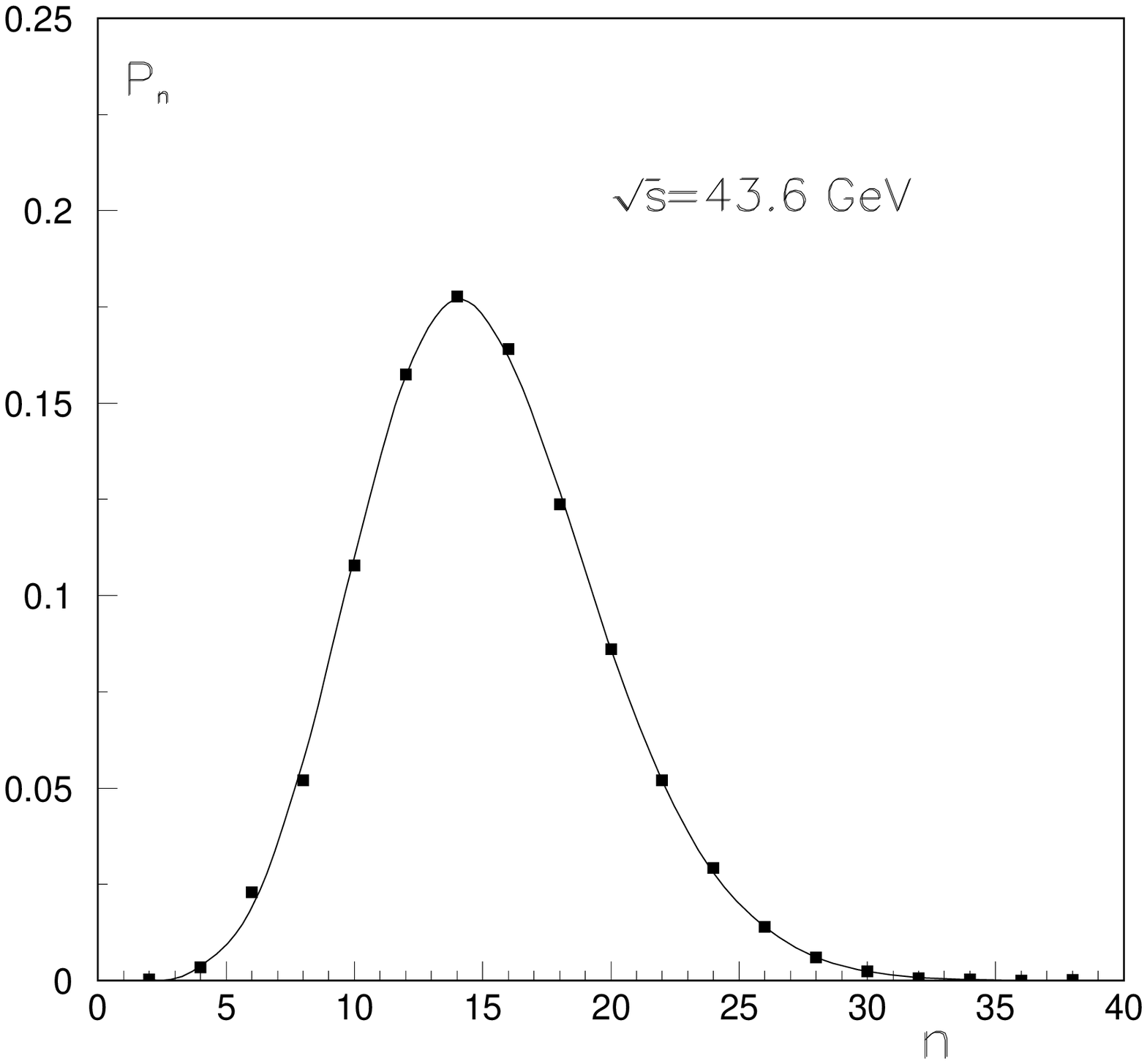}
\caption{MD at 43.6GeV.}
\label{4dfig}
\end{minipage}\hfill
\begin{minipage}[b]{.3\linewidth}
\centering
\includegraphics[width=\linewidth, height=2in, angle=0]{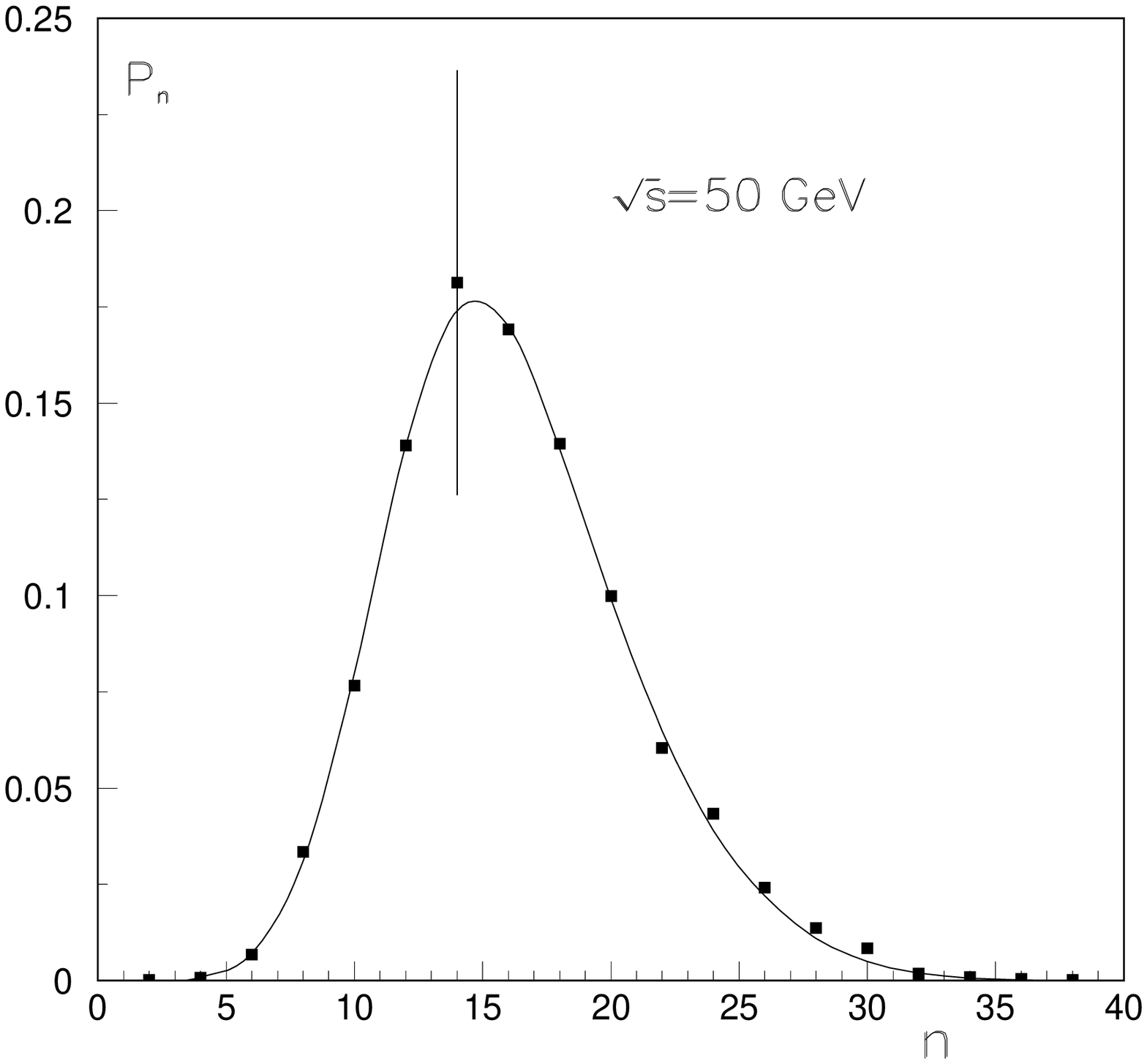}
\caption{MD at 50GeV.}
\label{5dfig}
\end{minipage}\hfill
\begin{minipage}[b]{.3\linewidth}
\centering
\includegraphics[width=\linewidth, height=2in, angle=0]{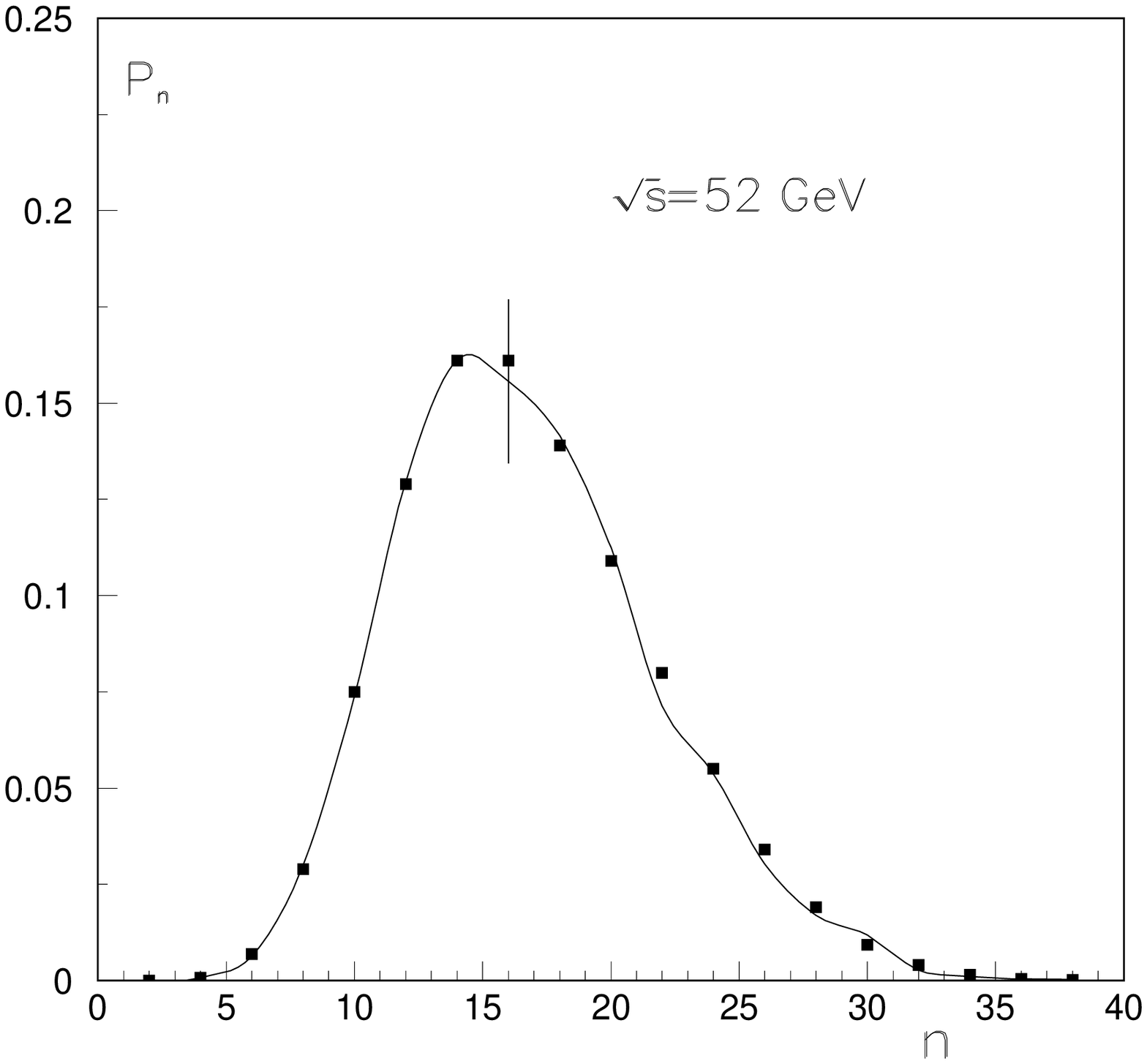}
\caption{MD at 52GeV.}
\label{6dfig}
\end{minipage}
\end{figure}

\begin{figure}
\begin{minipage}[b]{.3\linewidth}
\centering
\includegraphics[width=\linewidth, height=2in, angle=0]{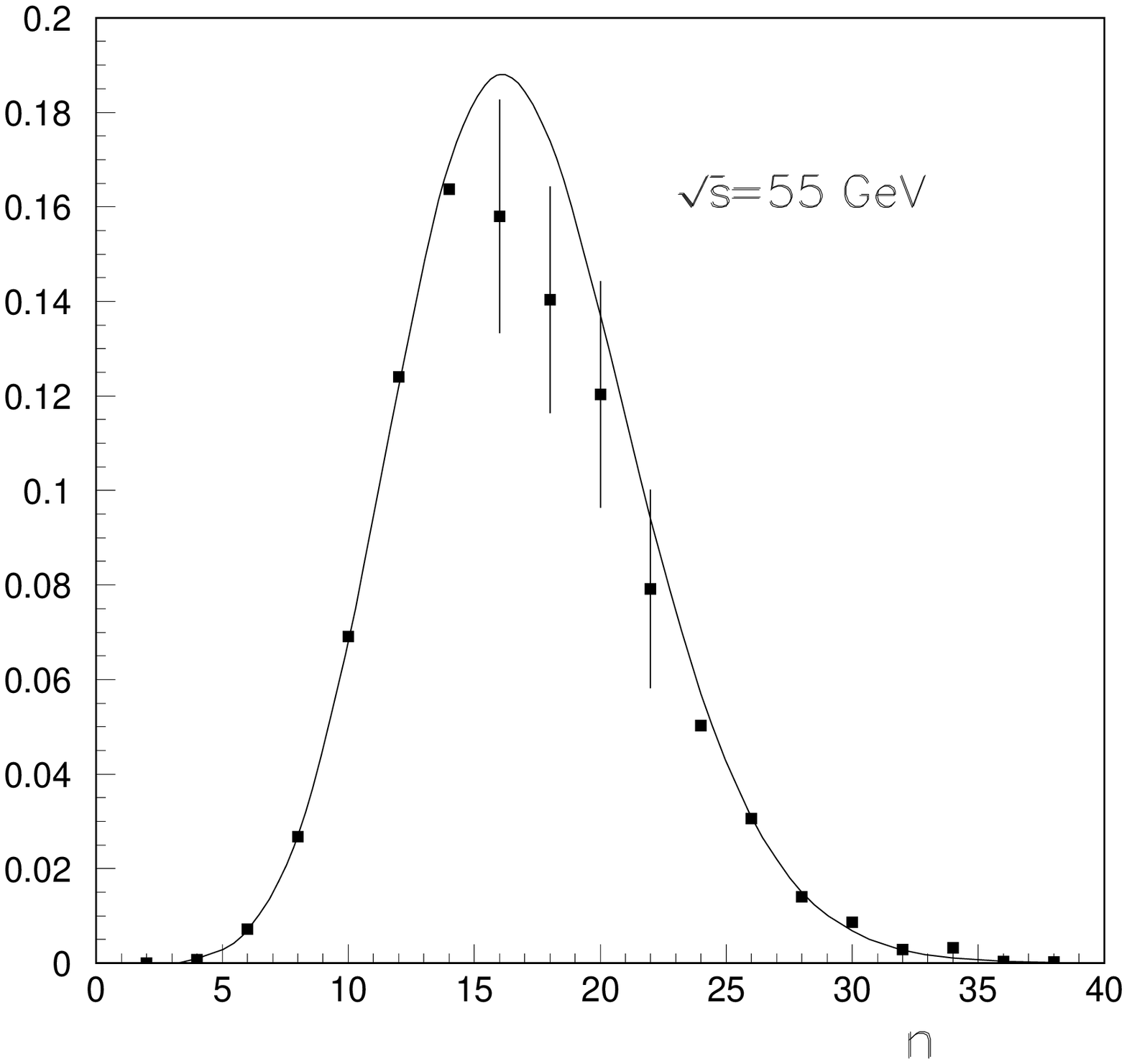}
\caption{MD at 55GeV.}
\label{7dfig}
\end{minipage}\hfill
\begin{minipage}[b]{.3\linewidth}
\centering
\includegraphics[width=\linewidth, height=2in, angle=0]{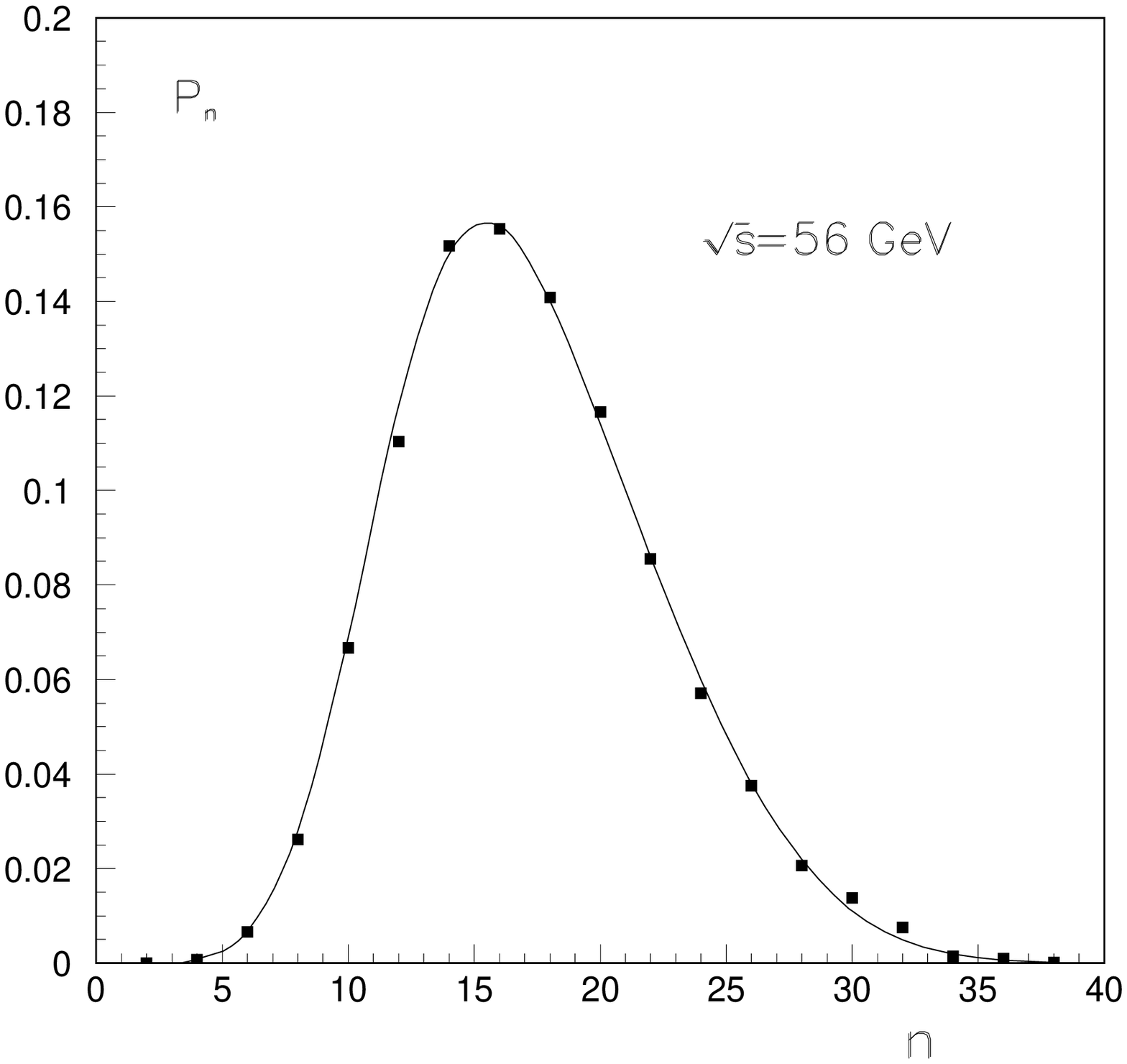}
\caption{MD at 56GeV.}
\label{8dfig}
\end{minipage}\hfill
\begin{minipage}[b]{.3\linewidth}
\centering
\includegraphics[width=\linewidth, height=2in, angle=0]{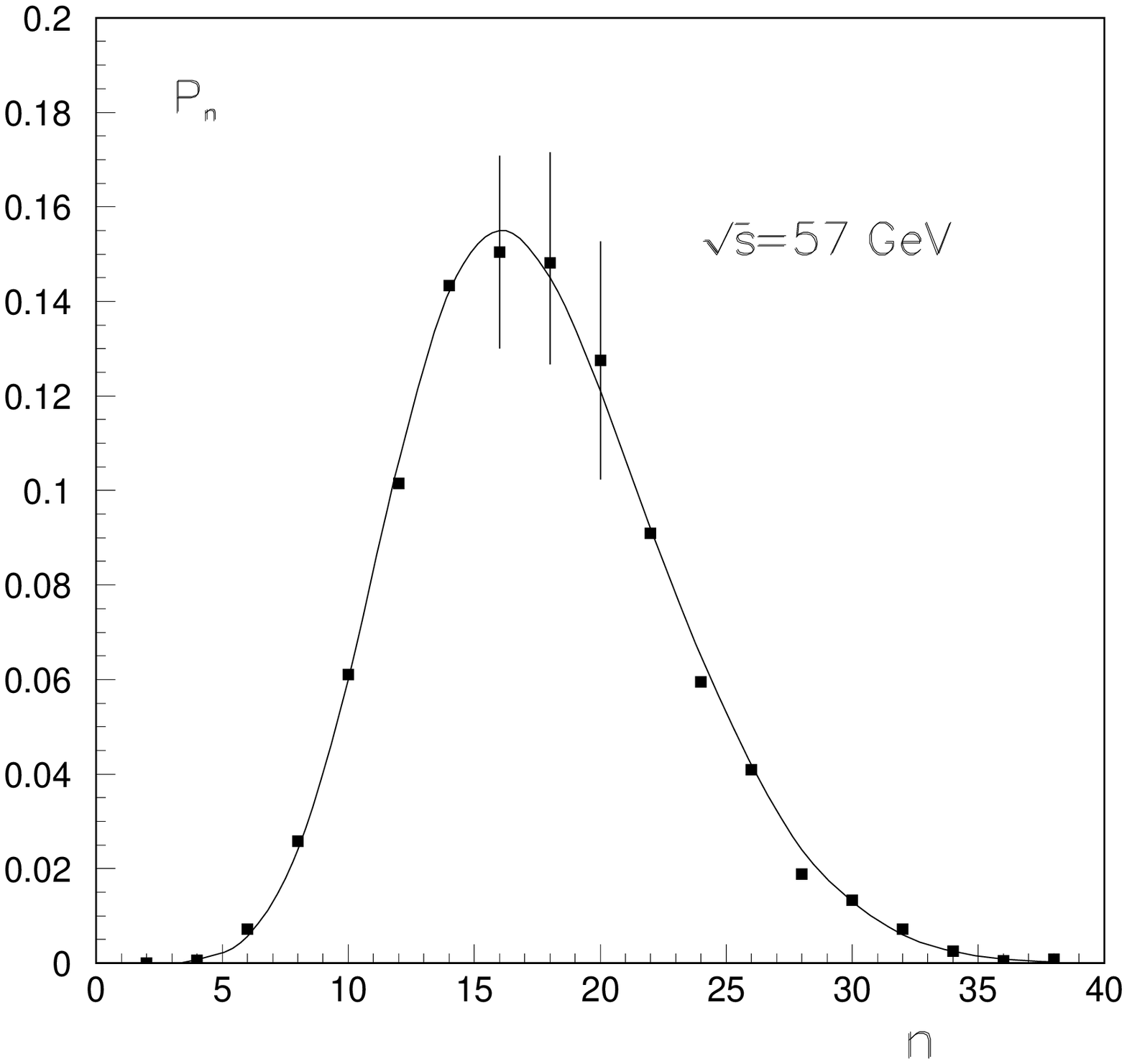}
\caption{MD at 57GeV.}
\label{9dfig}
\end{minipage}
\end{figure}

\begin{figure}
\begin{minipage}[b]{.3\linewidth}
\centering
\includegraphics[width=\linewidth, height=2in, angle=0]{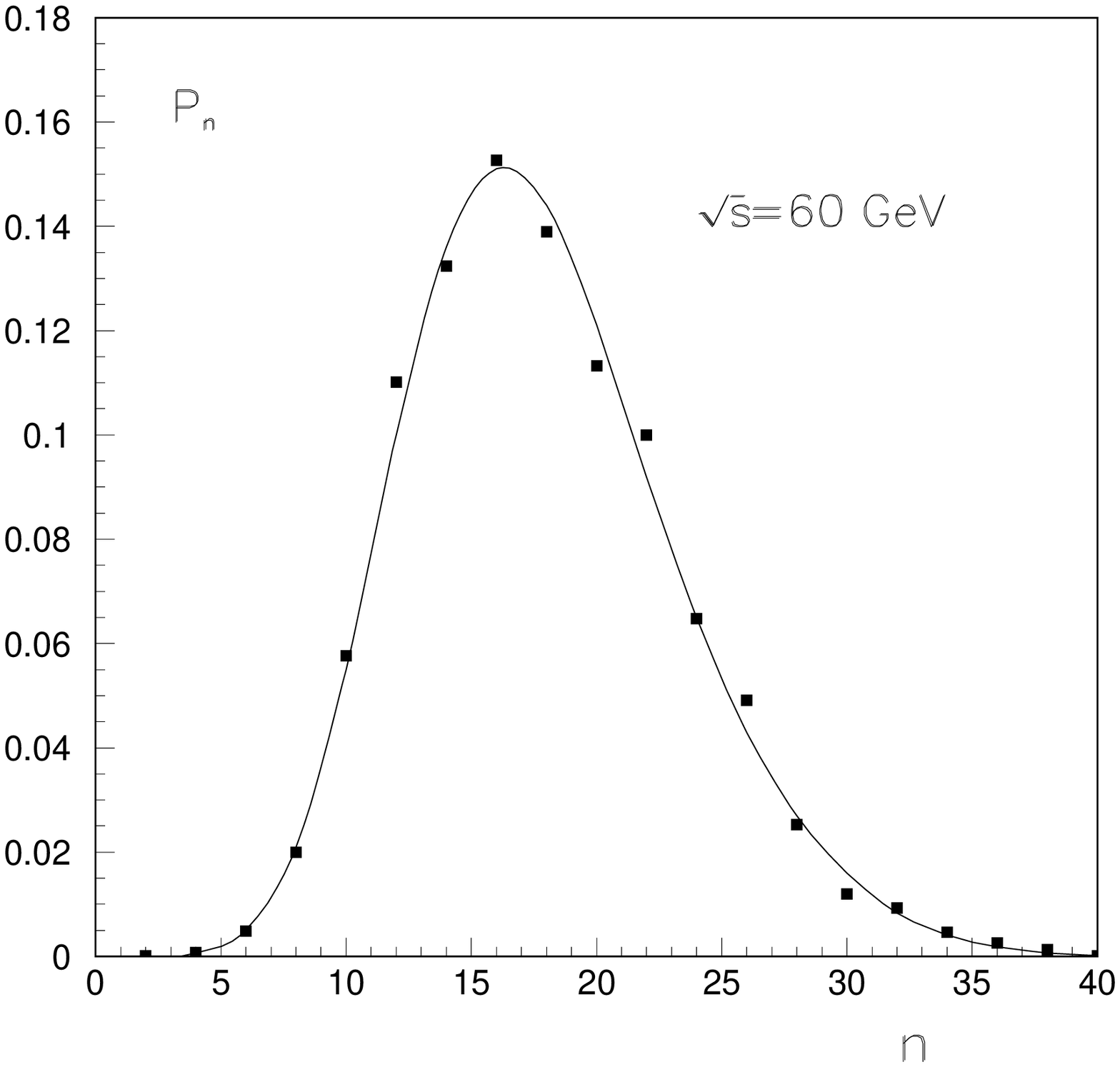}
\caption{MD at 60GeV.}
\label{10dfig}
\end{minipage}\hfill
\begin{minipage}[b]{.3\linewidth}
\centering
\includegraphics[width=\linewidth, height=2in, angle=0]{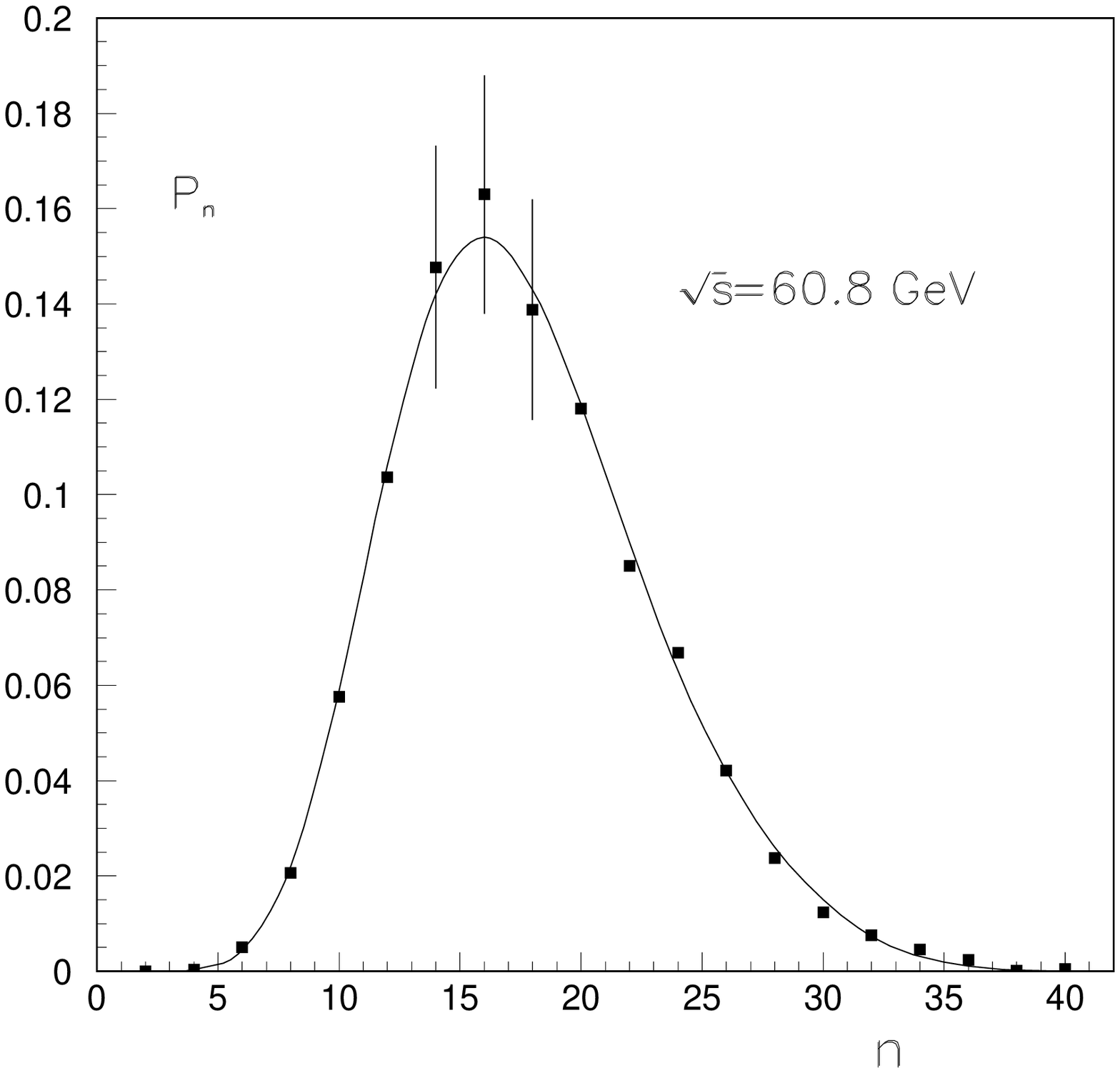}
\caption{MD at 60.8GeV.}
\label{11dfig}
\end{minipage}\hfill
\begin{minipage}[b]{.3\linewidth}
\centering
\includegraphics[width=\linewidth, height=2in, angle=0]{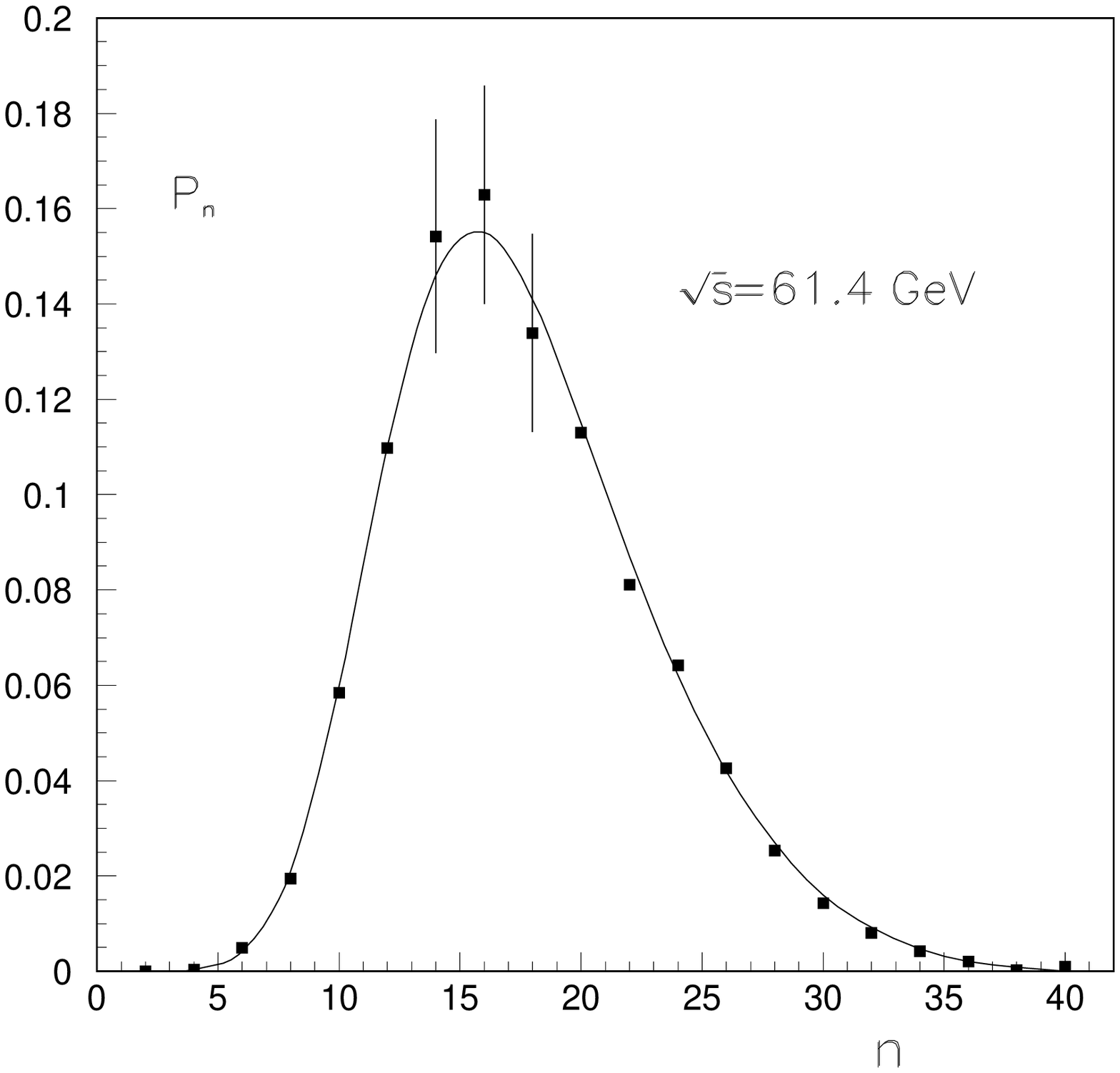}
\caption{MD at 61.4GeV.}
\label{12dfig}
\end{minipage}
\end{figure}

          \begin{figure}[htp]
\centerline{ \epsfig{file=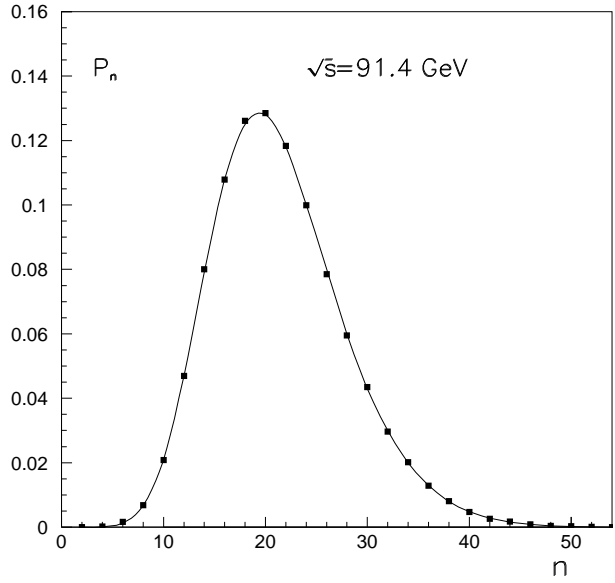,width=8.0cm}}
     \caption{MD at 91.4GeV. } \label{dddd}
     \label{13dfig}
          \end{figure}


\begin{figure}
\begin{minipage}[b]{.3\linewidth}
\centering
\includegraphics[width=\linewidth, height=2in, angle=0]{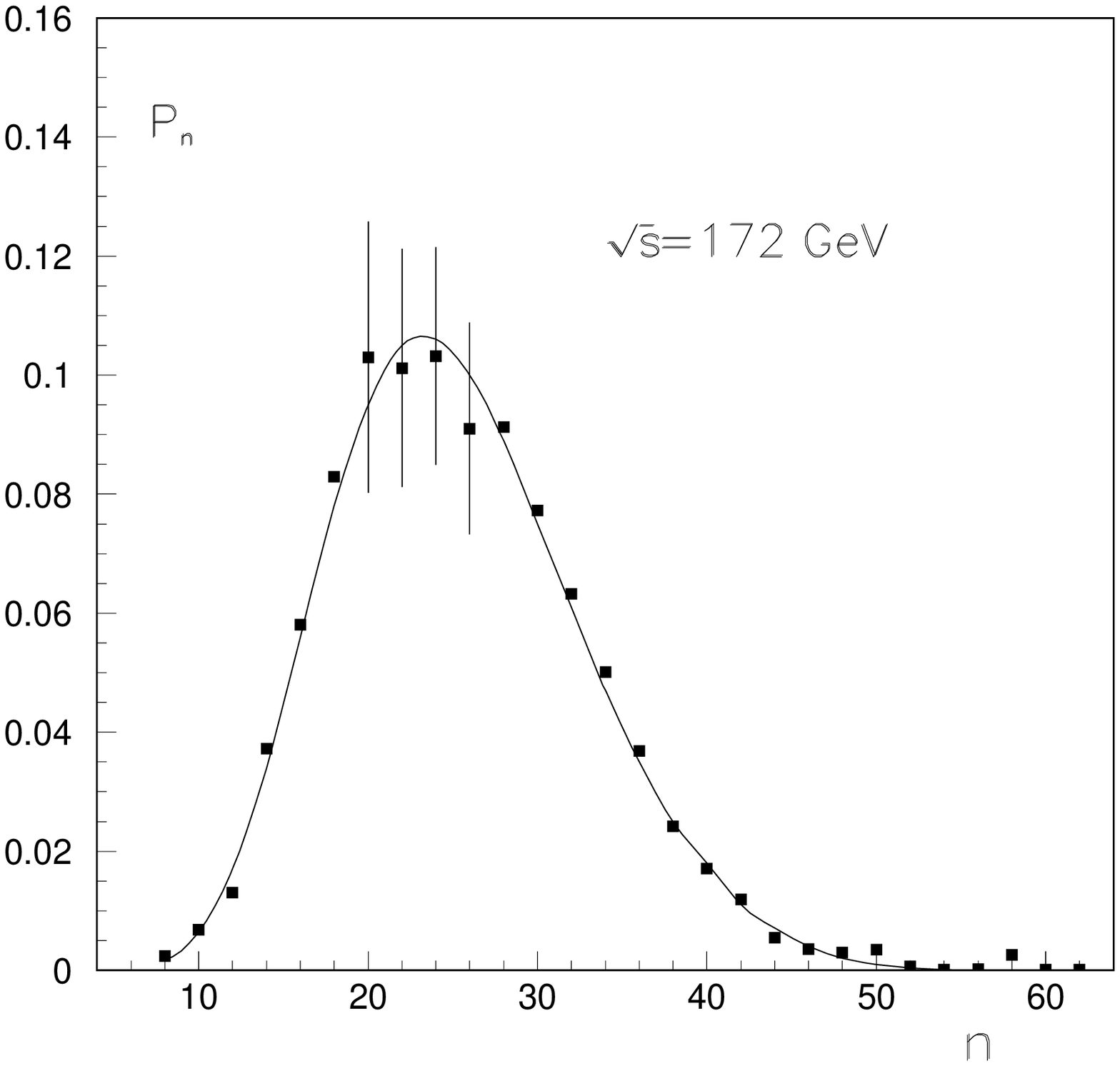}
\caption{MD at 172GeV.}
\label{14dfig}
\end{minipage}\hfill
\begin{minipage}[b]{.3\linewidth}
\centering
\includegraphics[width=\linewidth, height=2in, angle=0]{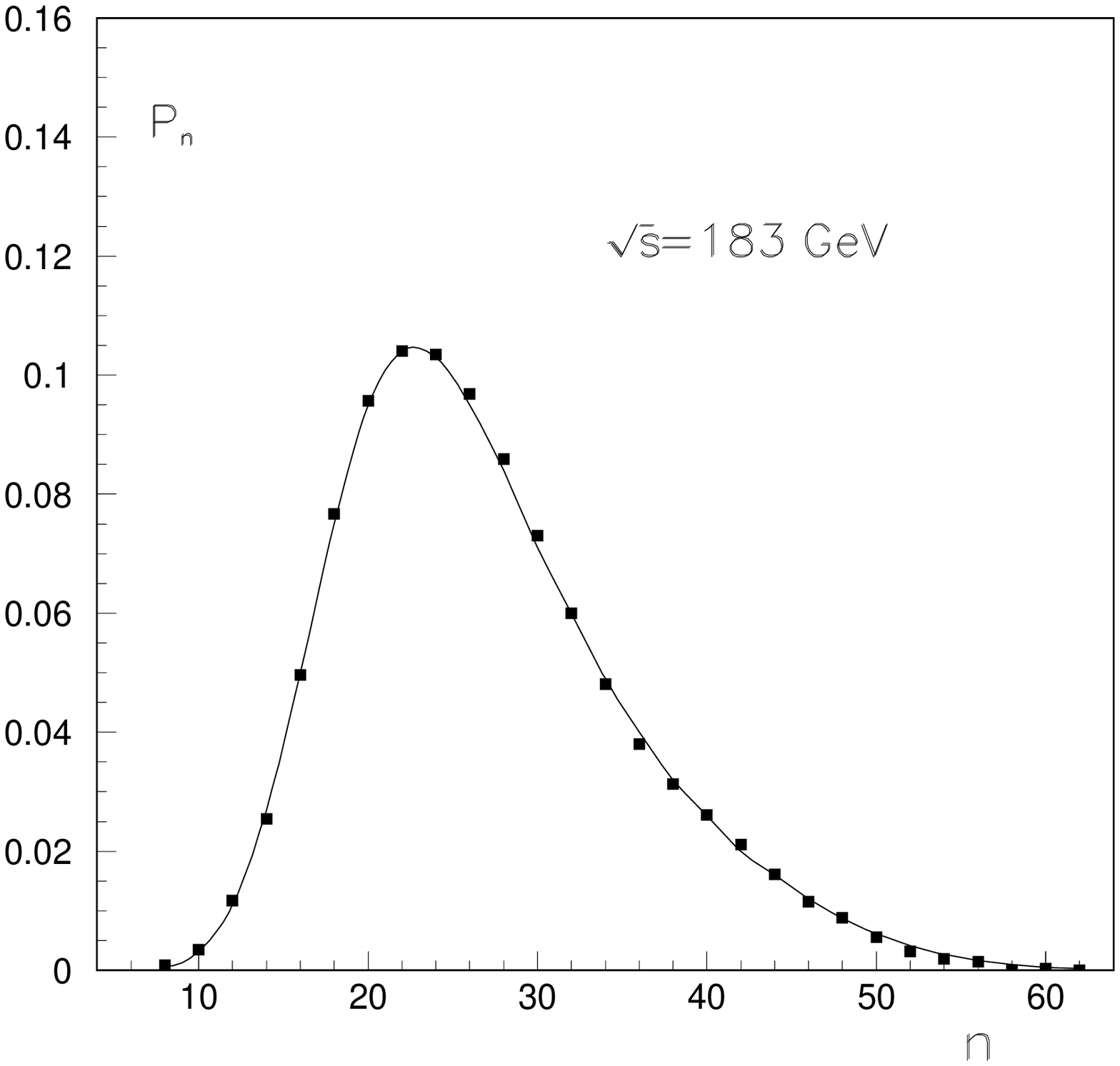}
\caption{MD at 183GeV.}
\label{15dfig}
\end{minipage}\hfill
\begin{minipage}[b]{.3\linewidth}
\centering
\includegraphics[width=\linewidth, height=2in, angle=0]{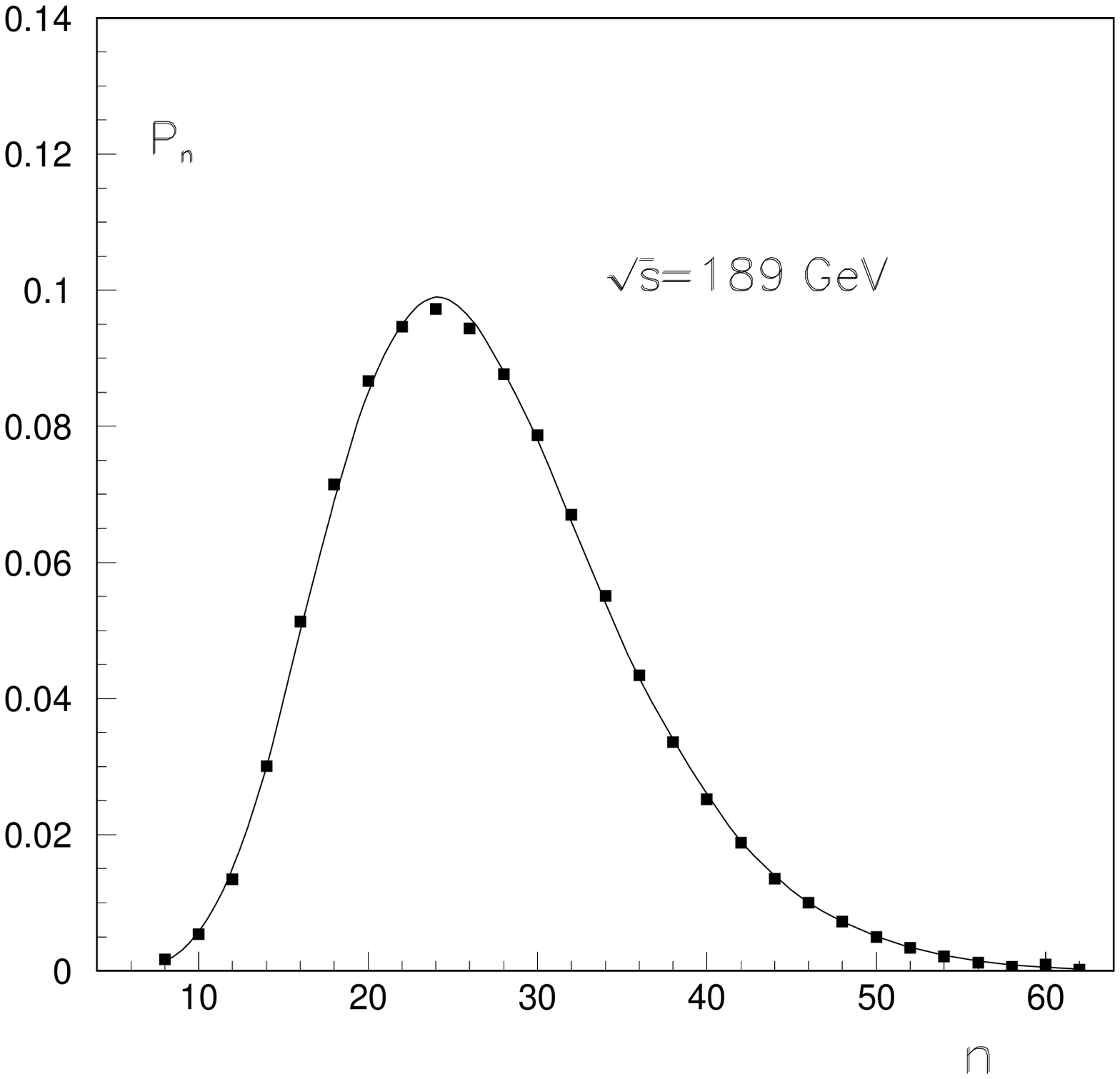}
\caption{MD at 189GeV.}
\label{16dfig}
\end{minipage}
\end{figure}

\begin{figure}[h!]
 \leavevmode
\begin{minipage}[b]{.475\linewidth}
\centering
\includegraphics[width=\linewidth, height=3in, angle=0]{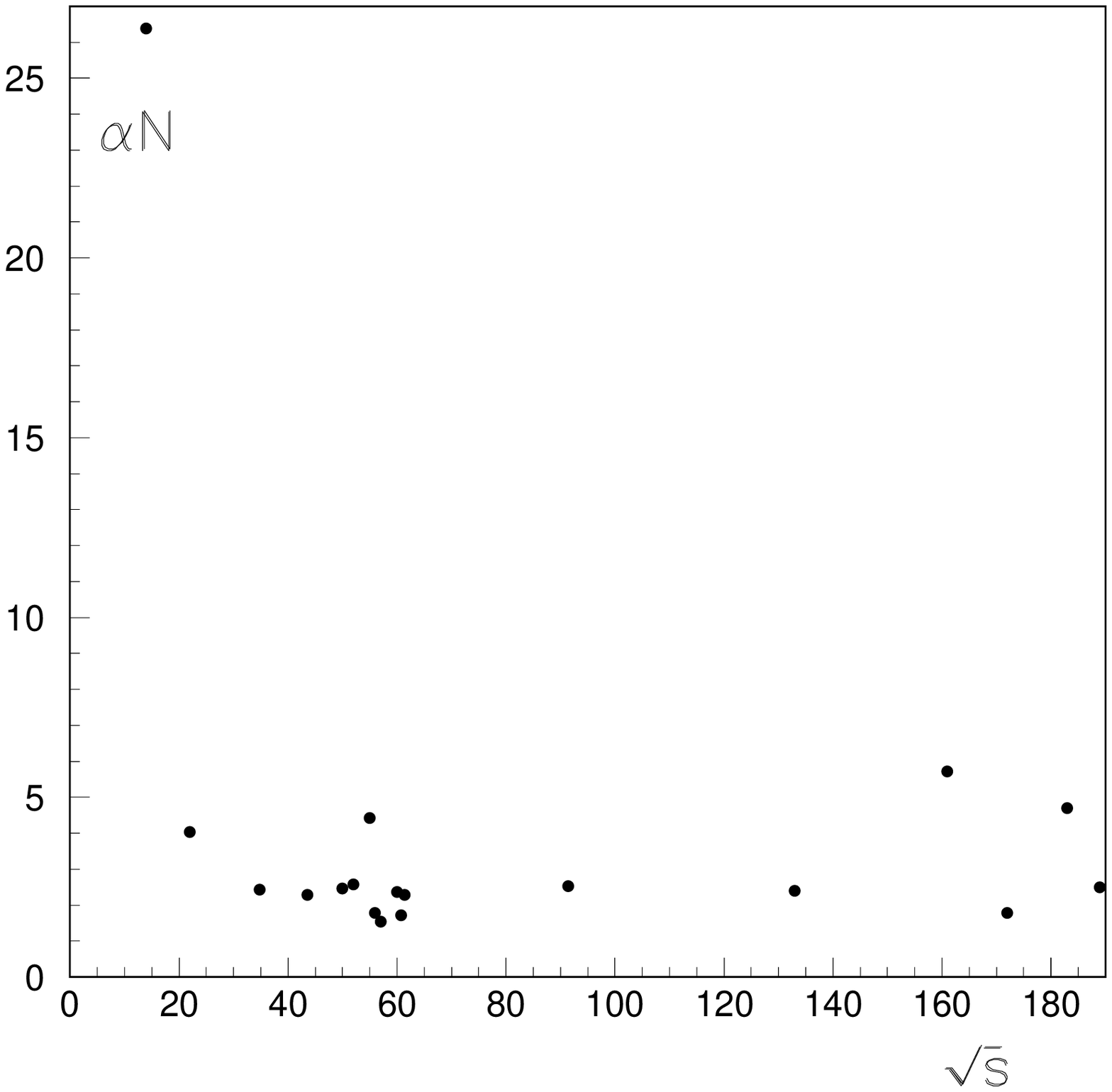}
\caption{Parameter $N_g=\alpha N_q$.}
\label{secondfig}
\end{minipage}\hfill
\begin{minipage}[b]{.475\linewidth}
\centering
\includegraphics[width=\linewidth, height=3in, angle=0]{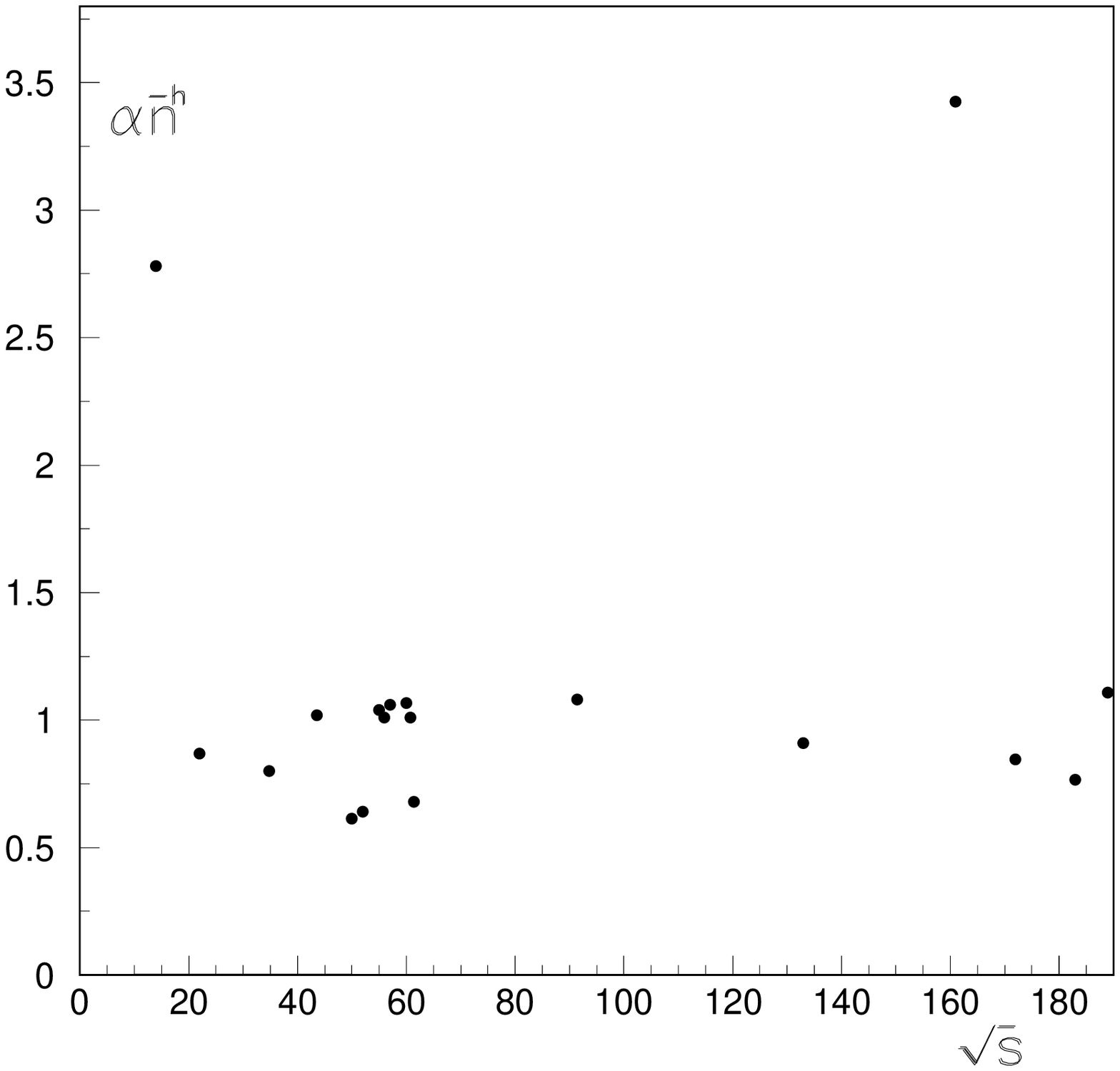}
\caption{Parameter $\overline n_g^h=\alpha \overline n_q^h$ .}
\label{threddfig}
\end{minipage}\\
\end{figure}
          \begin{figure}[htp]
\centerline{ \epsfig{file=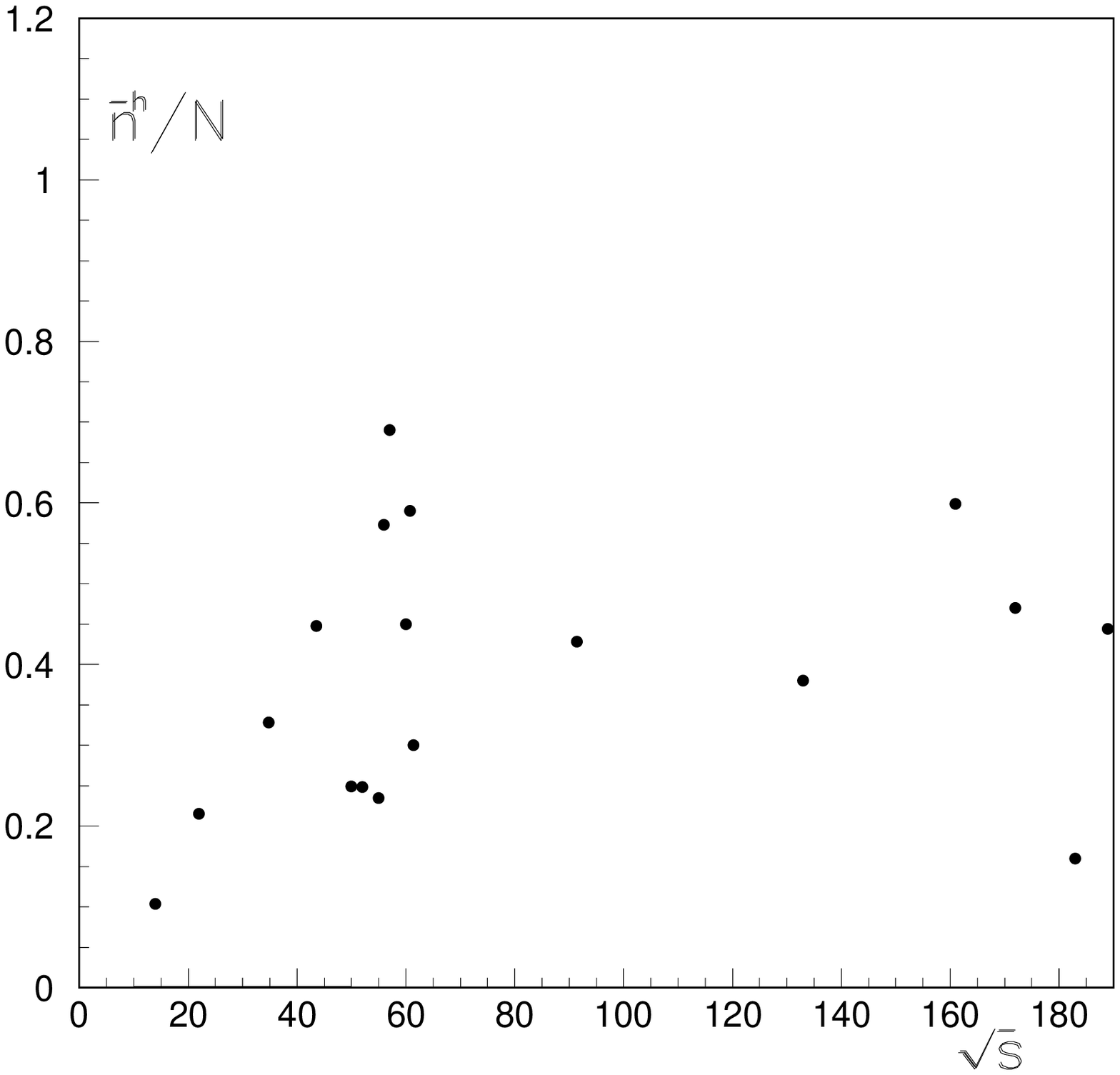,width=8.0cm}}
     \caption{Ratio $\frac{\overline n^h_q}{N_q}$. } \label{dddd}
          \end{figure}
\newpage
\begin{figure}
\begin{minipage}[b]{.3\linewidth}
\centering
\includegraphics[width=\linewidth, height=2in, angle=0]{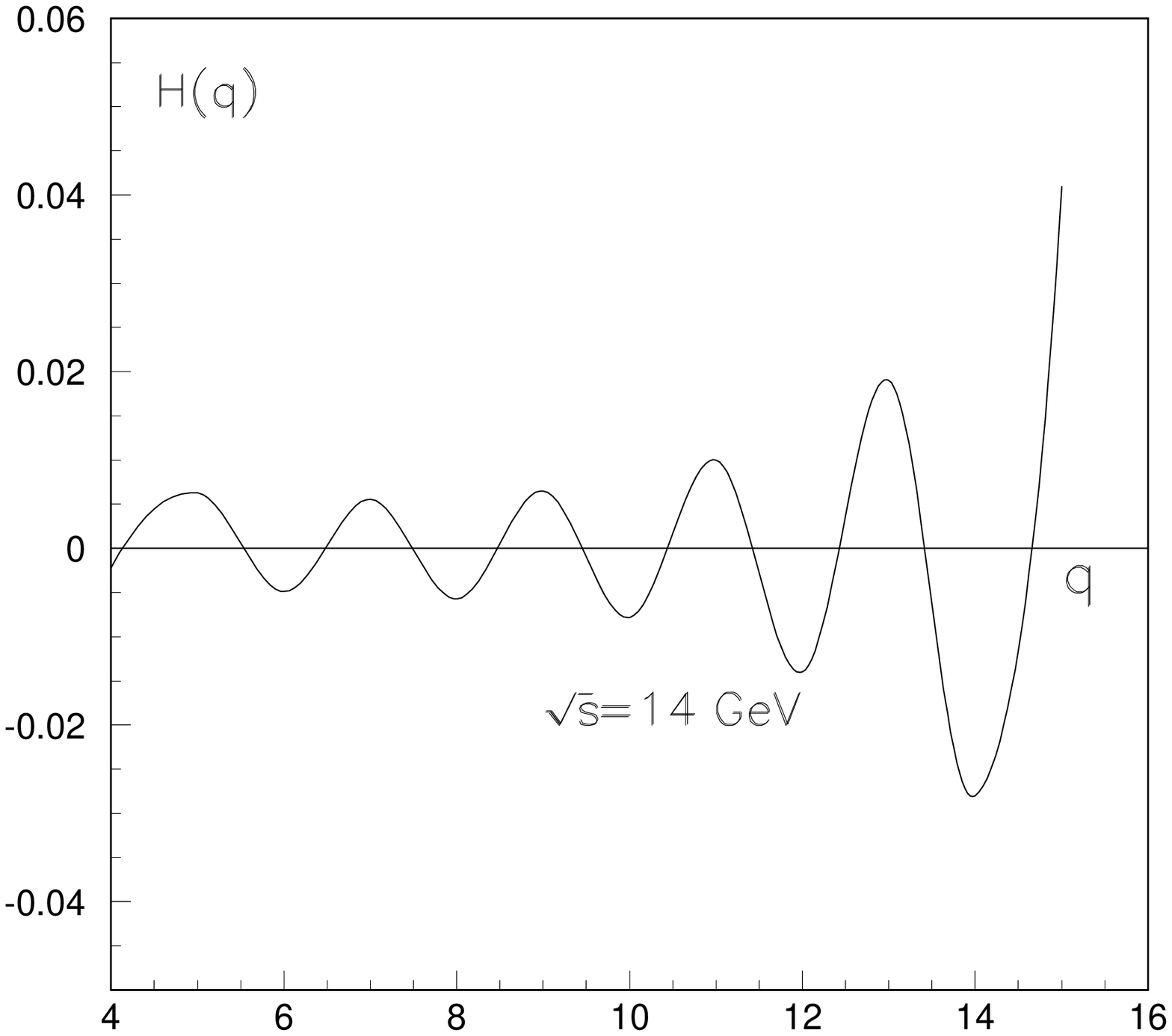}
\caption{$H_q$ at 14GeV.}
\label{1kdfig}
\end{minipage}\hfill
\begin{minipage}[b]{.3\linewidth}
\centering
\includegraphics[width=\linewidth, height=2in, angle=0]{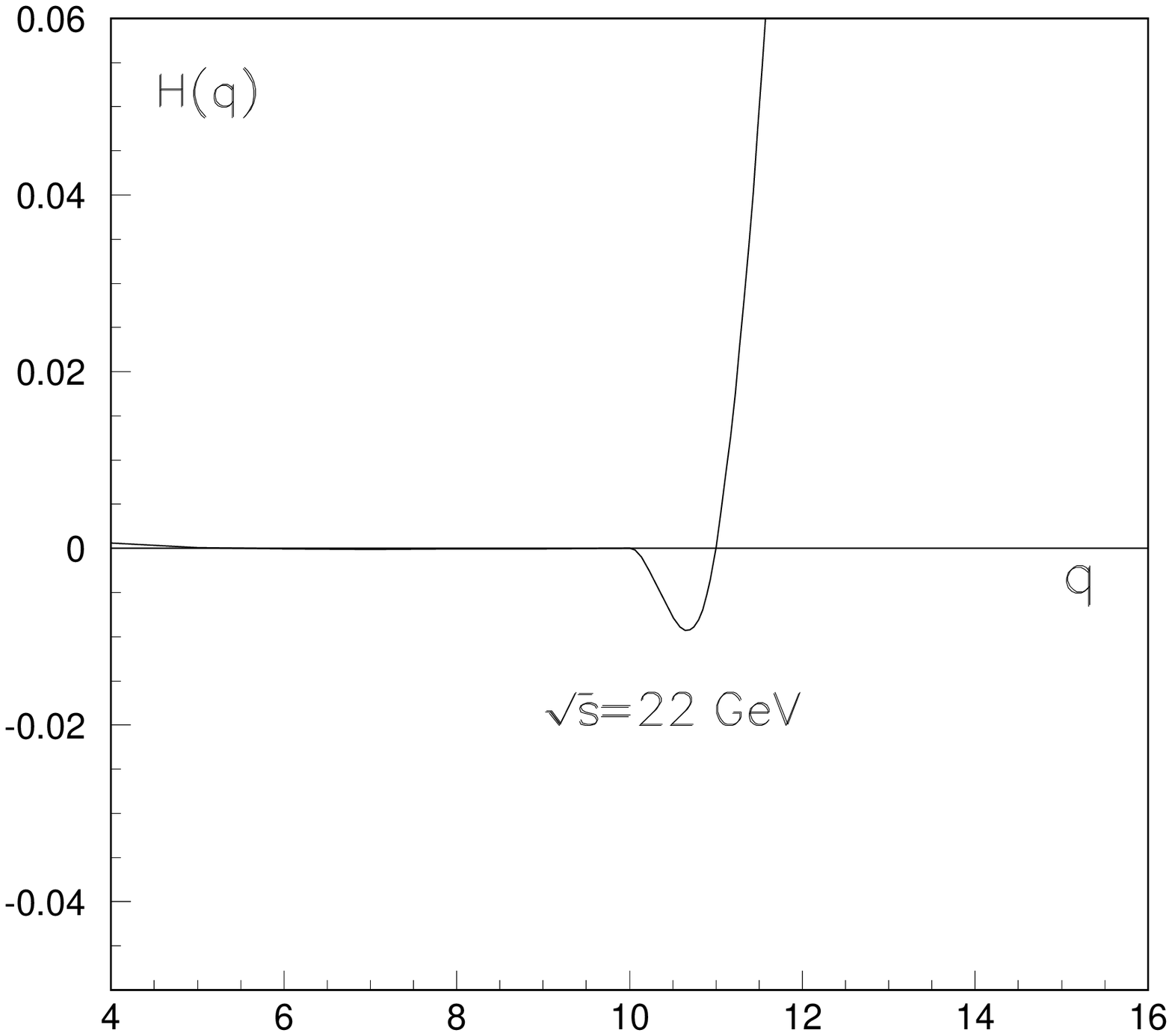}
\caption{$H_q$ at 22GeV.}
\label{2kdfig}
\end{minipage}\hfill
\begin{minipage}[b]{.3\linewidth}
\centering
\includegraphics[width=\linewidth, height=2in, angle=0]{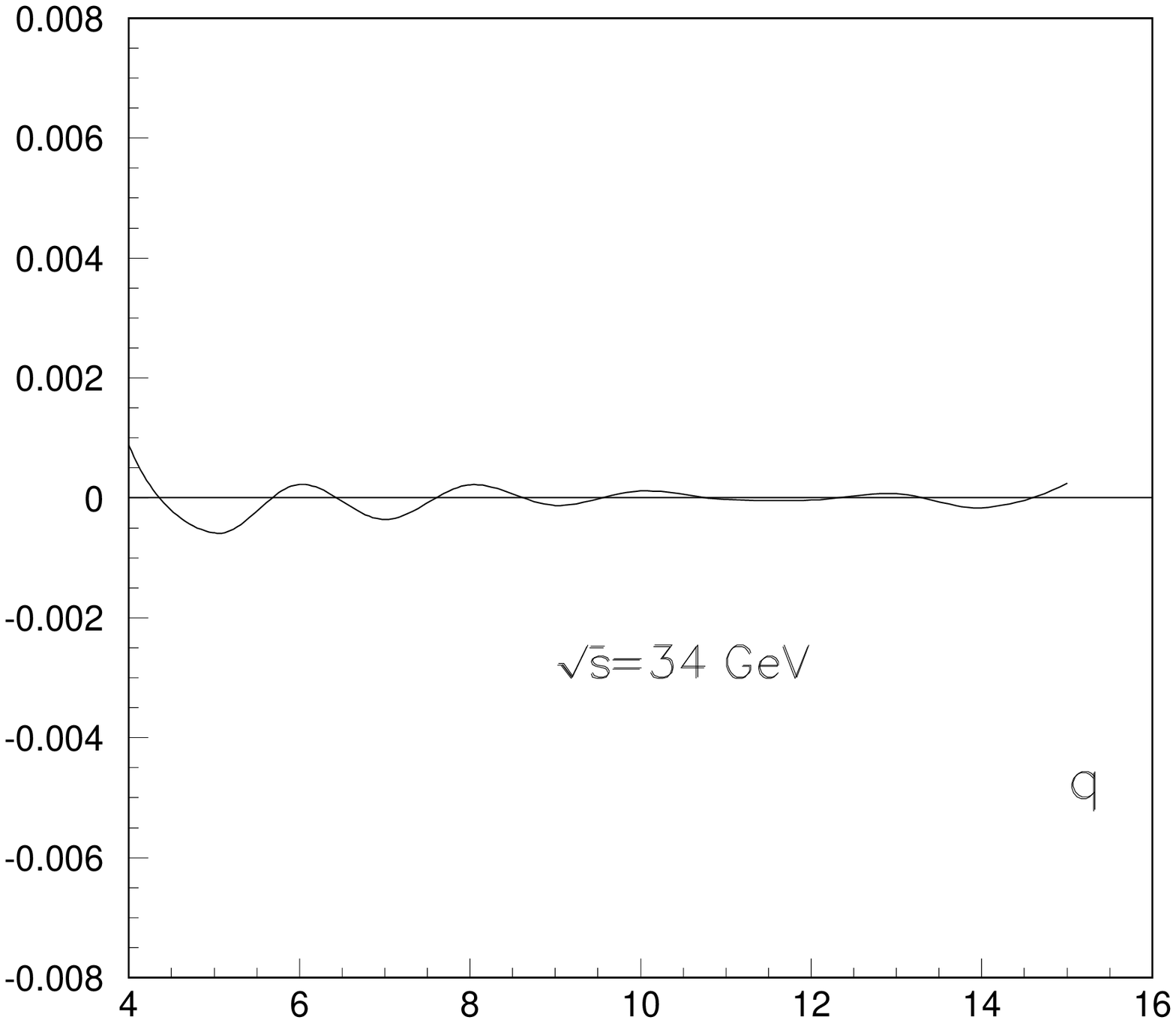}
\caption{ $H_q$ at 34.8GeV}
\label{3kdfig}
\end{minipage}
\end{figure}

\begin{figure}
\begin{minipage}[b]{.3\linewidth}
\centering
\includegraphics[width=\linewidth, height=2in, angle=0]{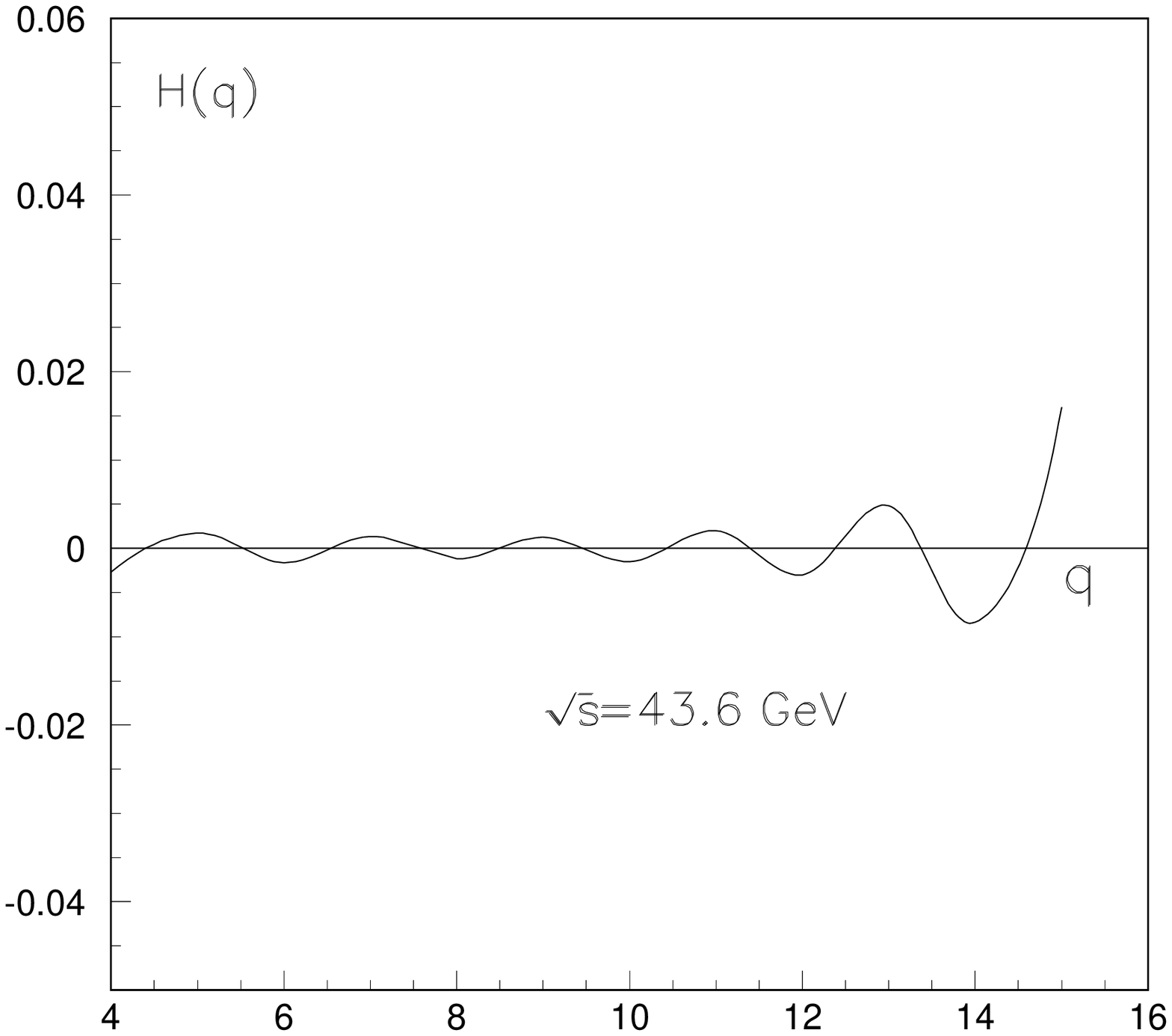}
\caption{$H_q$ at 43.6GeV.}
\label{4kdfig}
\end{minipage}\hfill
\begin{minipage}[b]{.3\linewidth}
\centering
\includegraphics[width=\linewidth, height=2in, angle=0]{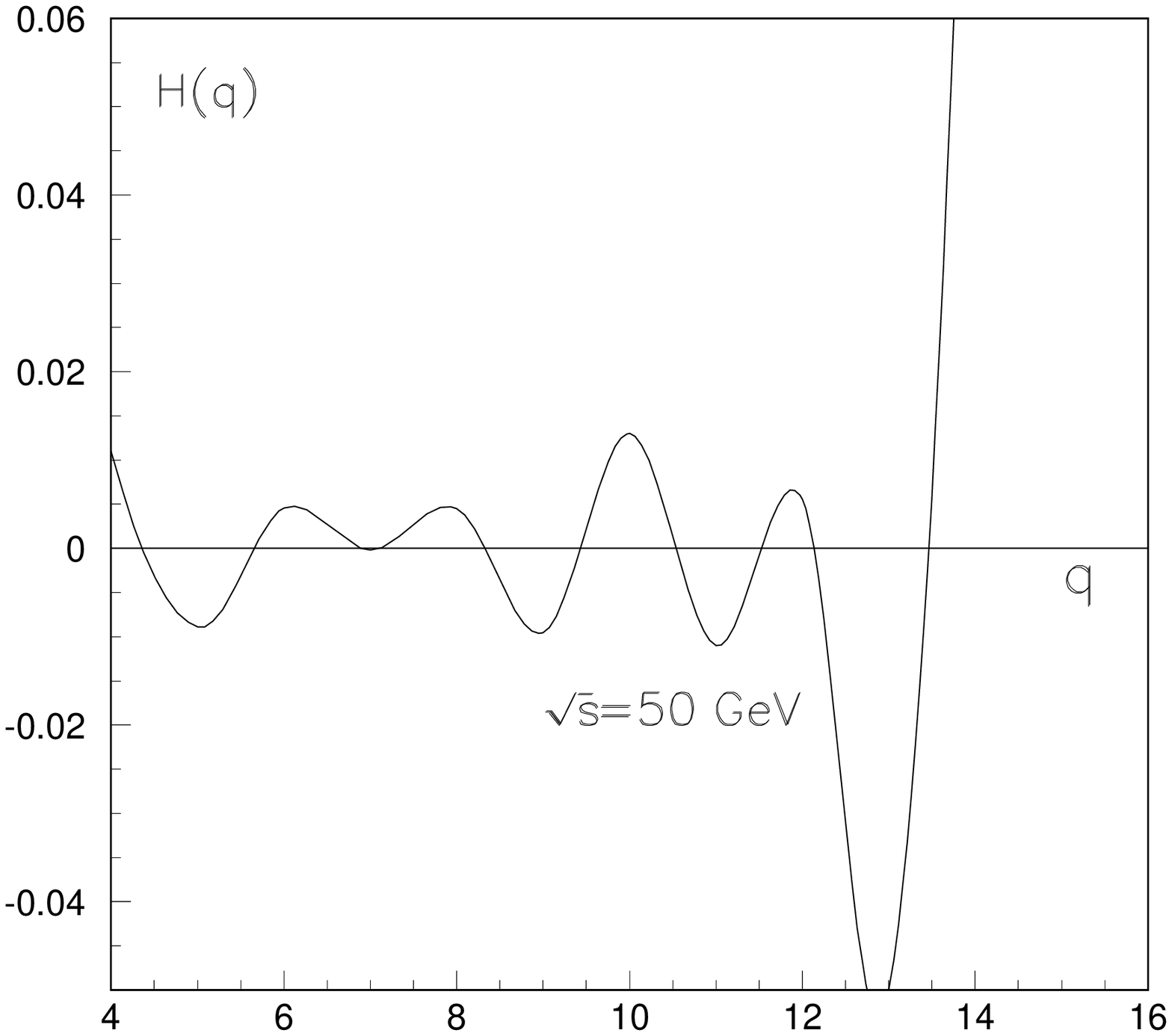}
\caption{$H_q$ at 50GeV.}
\label{5kdfig}
\end{minipage}\hfill
\begin{minipage}[b]{.3\linewidth}
\centering
\includegraphics[width=\linewidth, height=2in, angle=0]{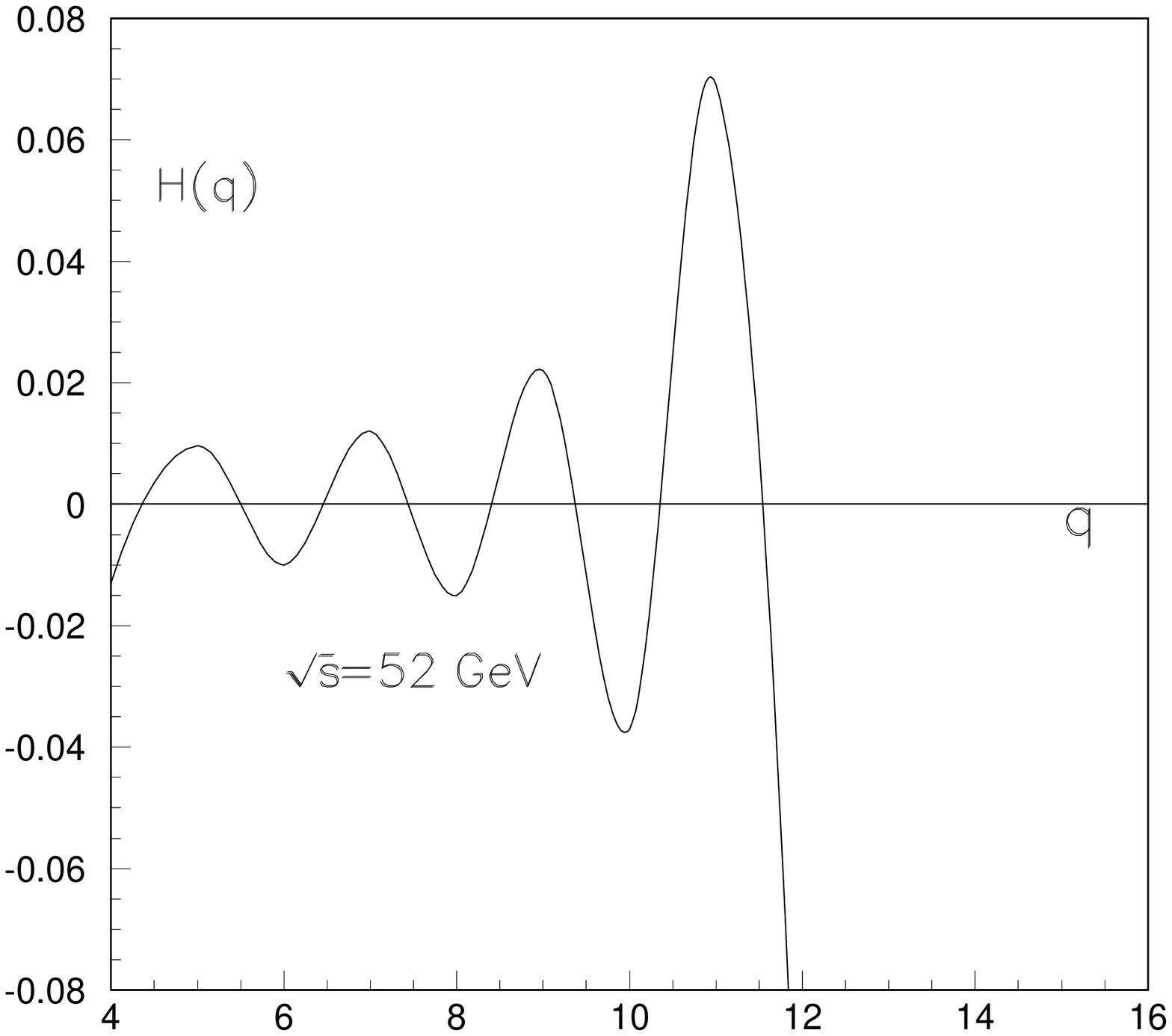}
\caption{$H_q$ at 52GeV.}
\label{6kdfig}
\end{minipage}
\end{figure}

\begin{figure}
\begin{minipage}[b]{.3\linewidth}
\centering
\includegraphics[width=\linewidth, height=2in, angle=0]{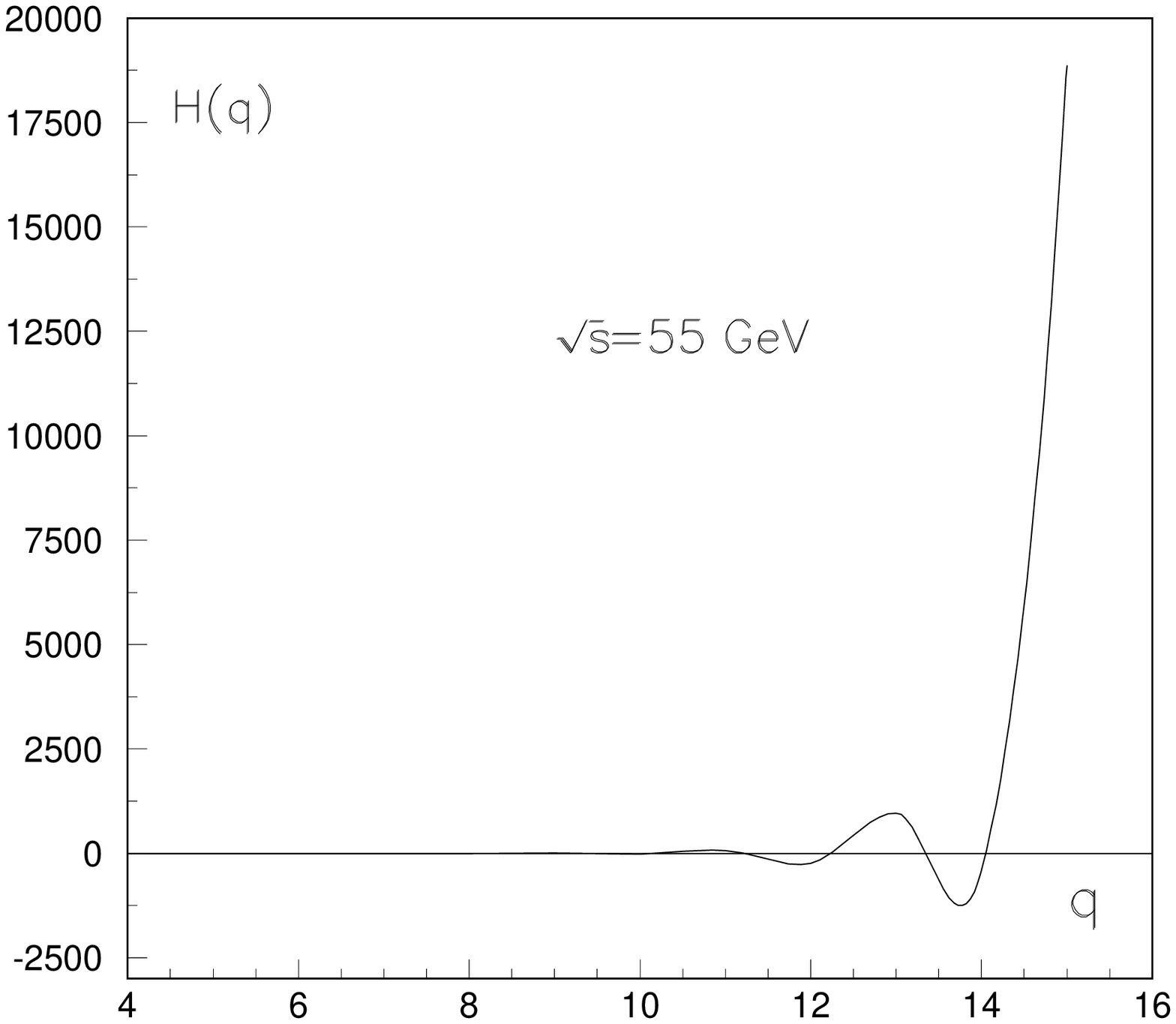}
\caption{$H_q$ at 55GeV.}
\label{7kdfig}
\end{minipage}\hfill
\begin{minipage}[b]{.3\linewidth}
\centering
\includegraphics[width=\linewidth, height=2in, angle=0]{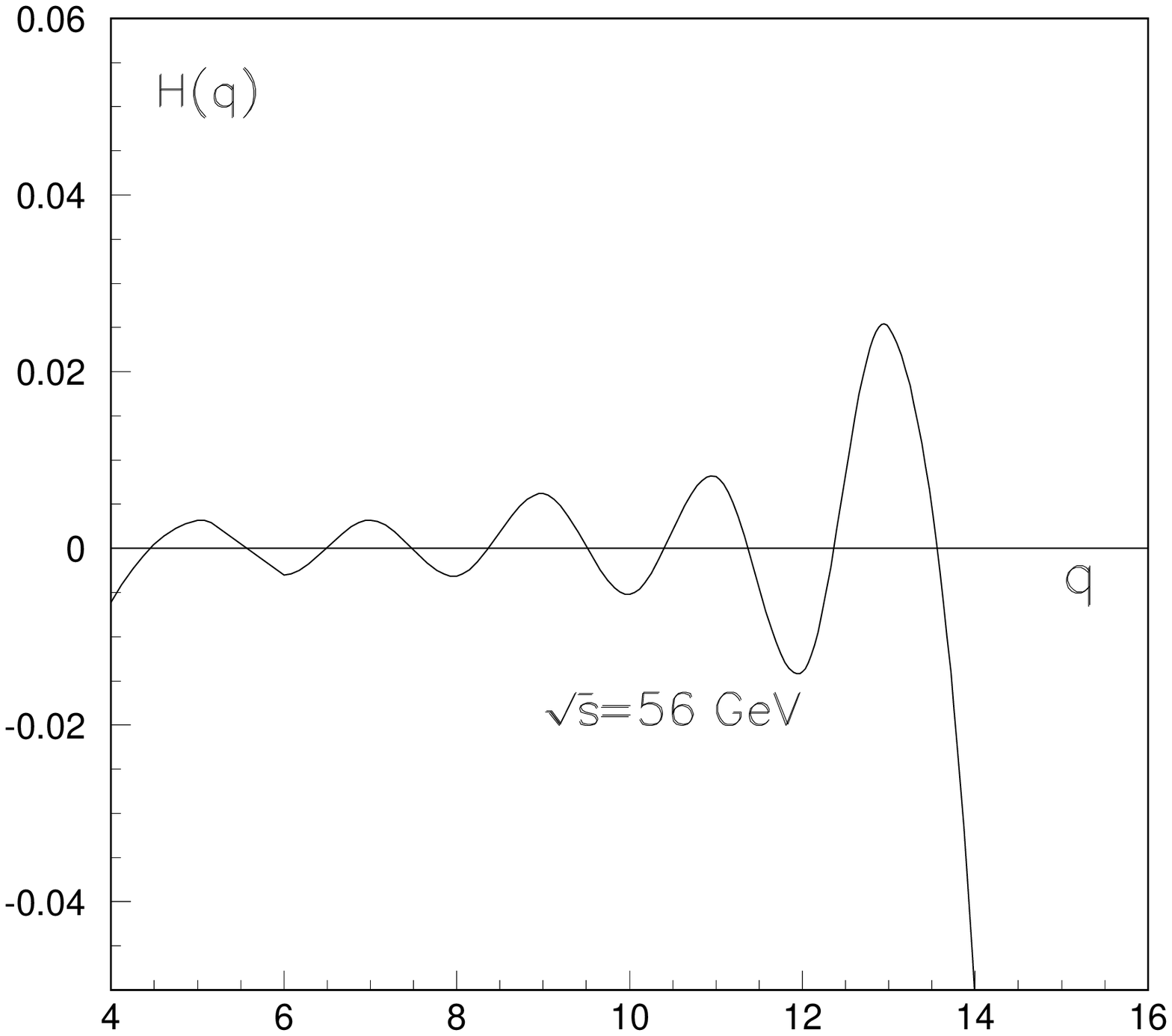}
\caption{$H_q$ at 56GeV.}
\label{8kdfig}
\end{minipage}\hfill
\begin{minipage}[b]{.3\linewidth}
\centering
\includegraphics[width=\linewidth, height=2in, angle=0]{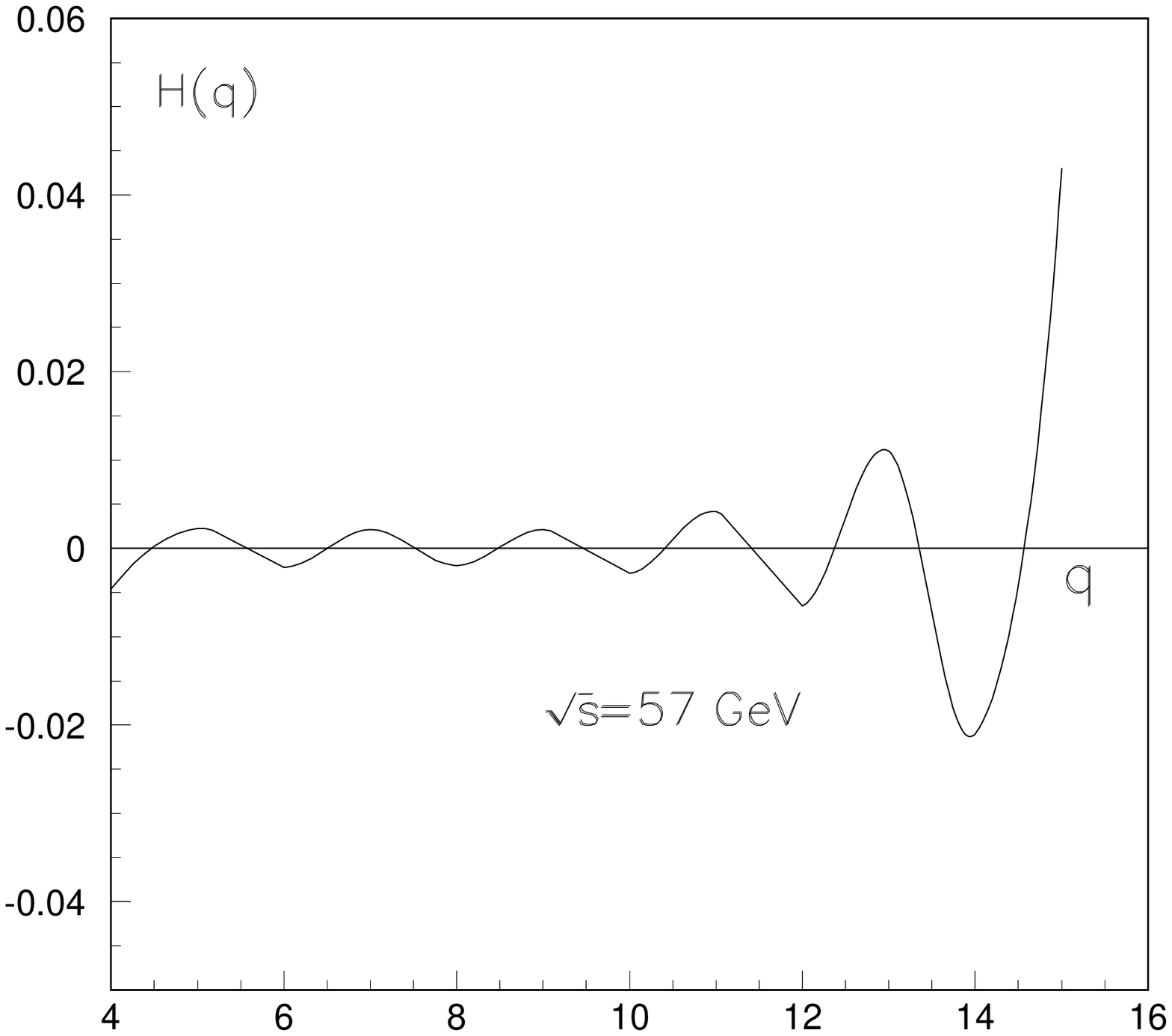}
\caption{$H_q$ at 57GeV.}
\label{9kdfig}
\end{minipage}
\end{figure}

\begin{figure}
\begin{minipage}[b]{.3\linewidth}
\centering
\includegraphics[width=\linewidth, height=2in, angle=0]{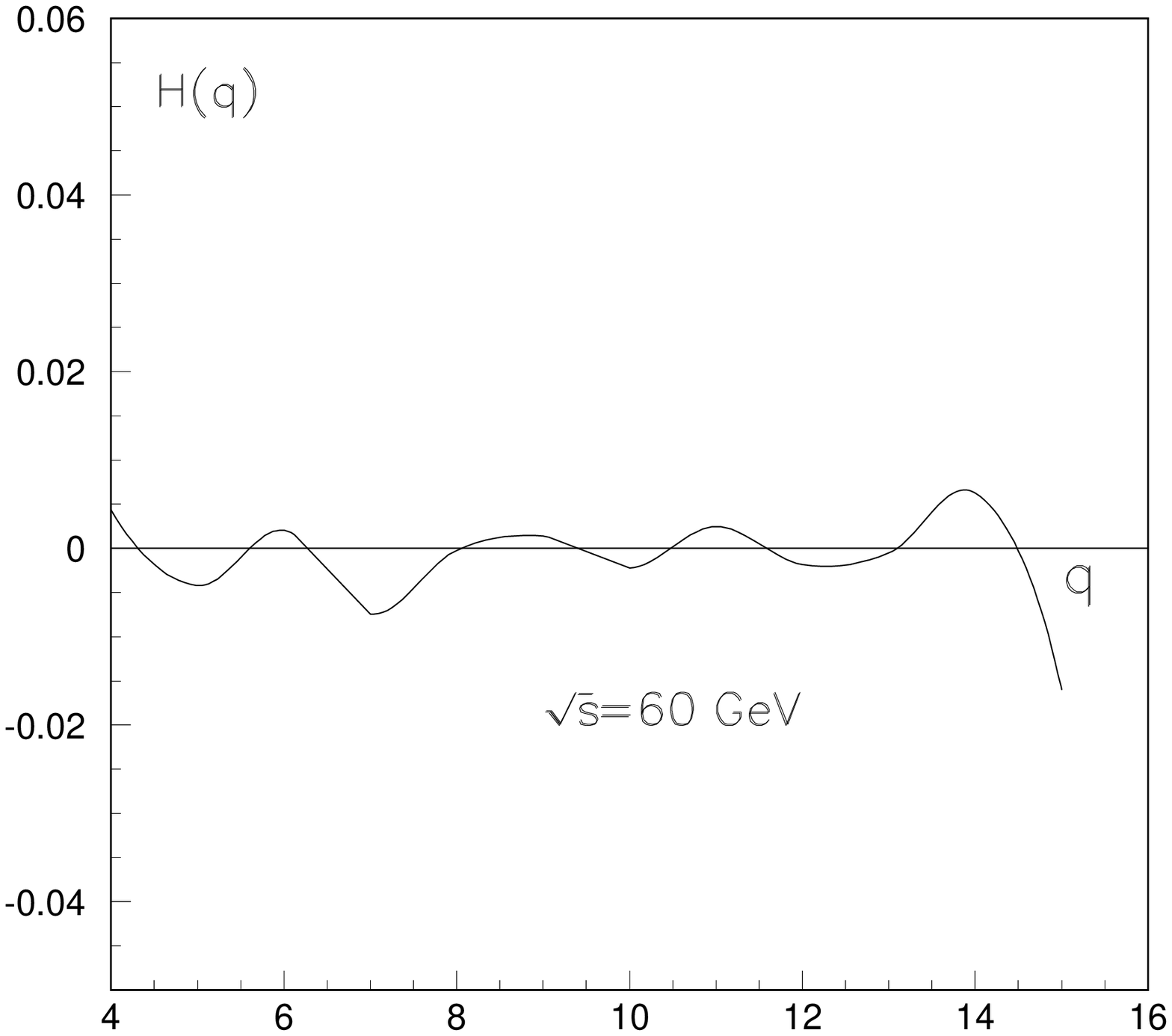}
\caption{$H_q$ at 60GeV.}
\label{10kdfig}
\end{minipage}\hfill
\begin{minipage}[b]{.3\linewidth}
\centering
\includegraphics[width=\linewidth, height=2in, angle=0]{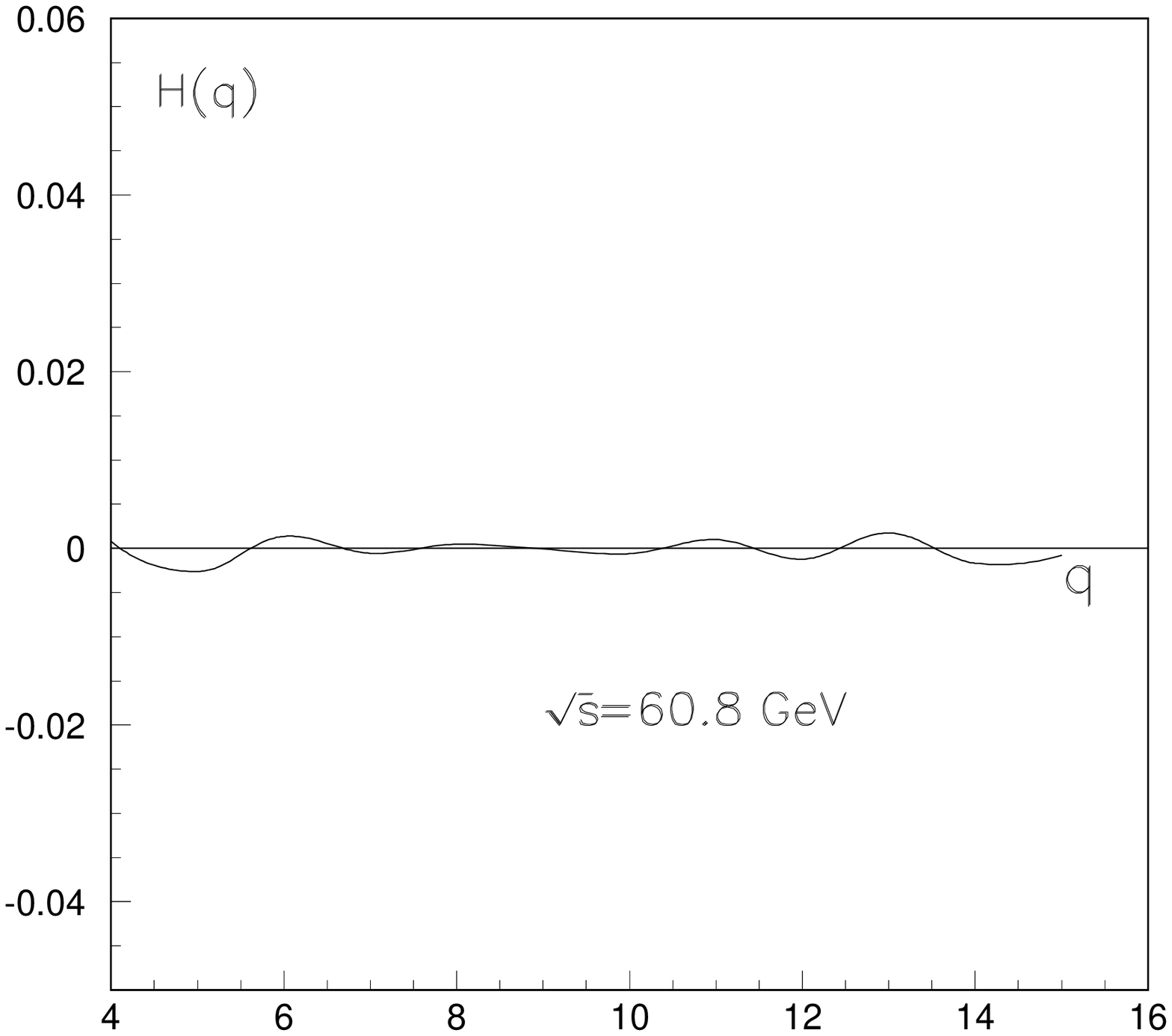}
\caption{$H_q$ at 60.8GeV.}
\label{11kdfig}
\end{minipage}\hfill
\begin{minipage}[b]{.3\linewidth}
\centering
\includegraphics[width=\linewidth, height=2in, angle=0]{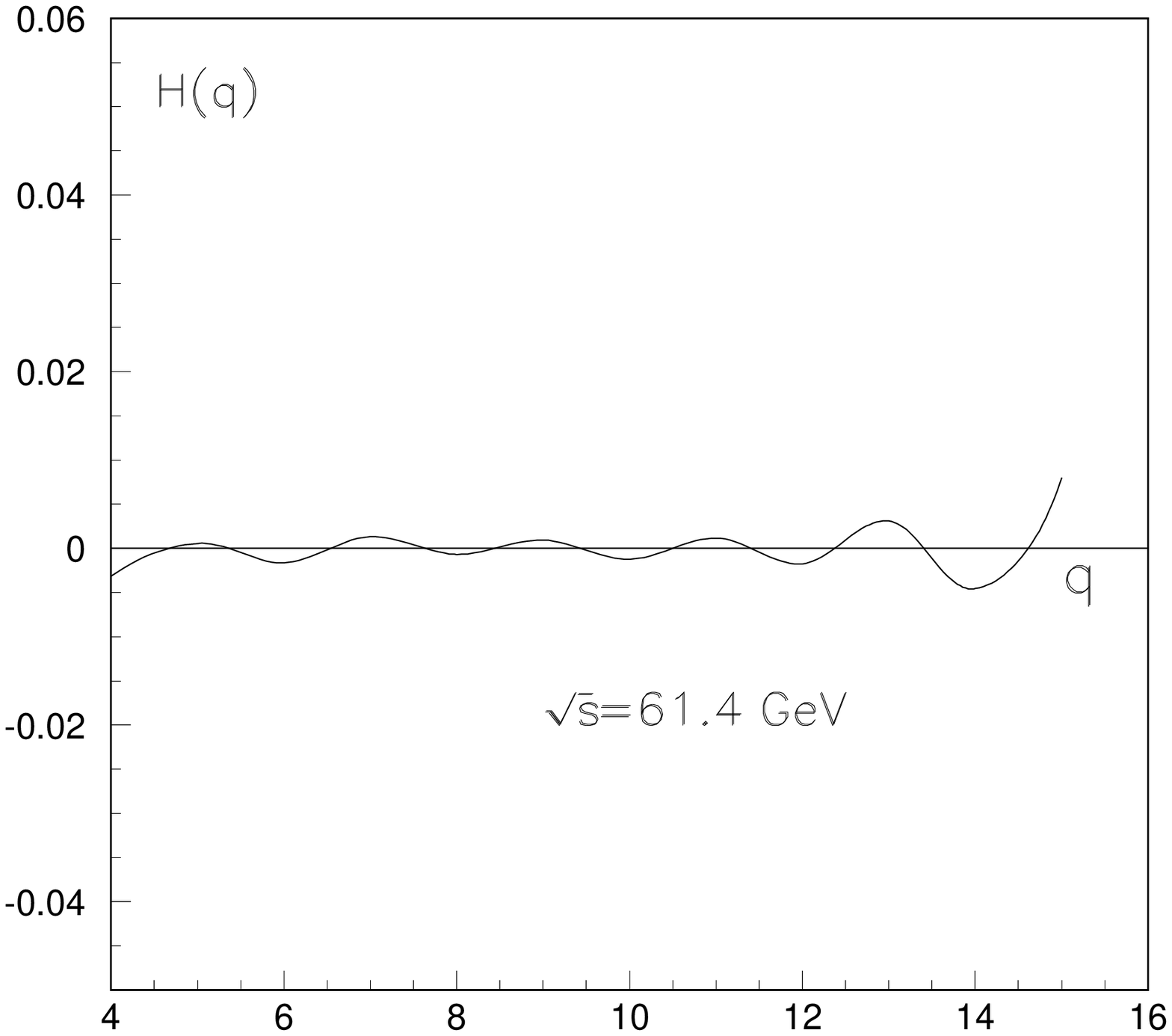}
\caption{$H_q$ at 61.4GeV.}
\label{12kdfig}
\end{minipage}
\end{figure}

\begin{figure}[htp]
\centerline{ \epsfig{file=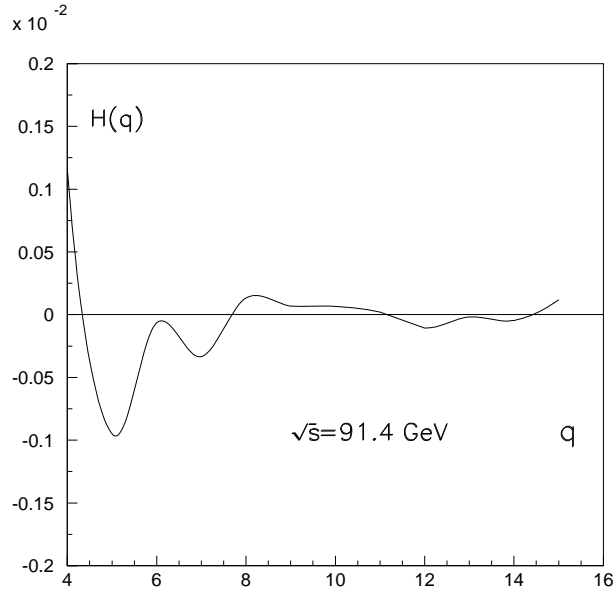,width=8.0cm}}
\caption{$H_q$ at 91.4GeV.} \label{kkkk}
\label{13kdfig}
\end{figure}

\begin{figure}
\begin{minipage}[b]{.3\linewidth}
\centering
\includegraphics[width=\linewidth, height=2in, angle=0]{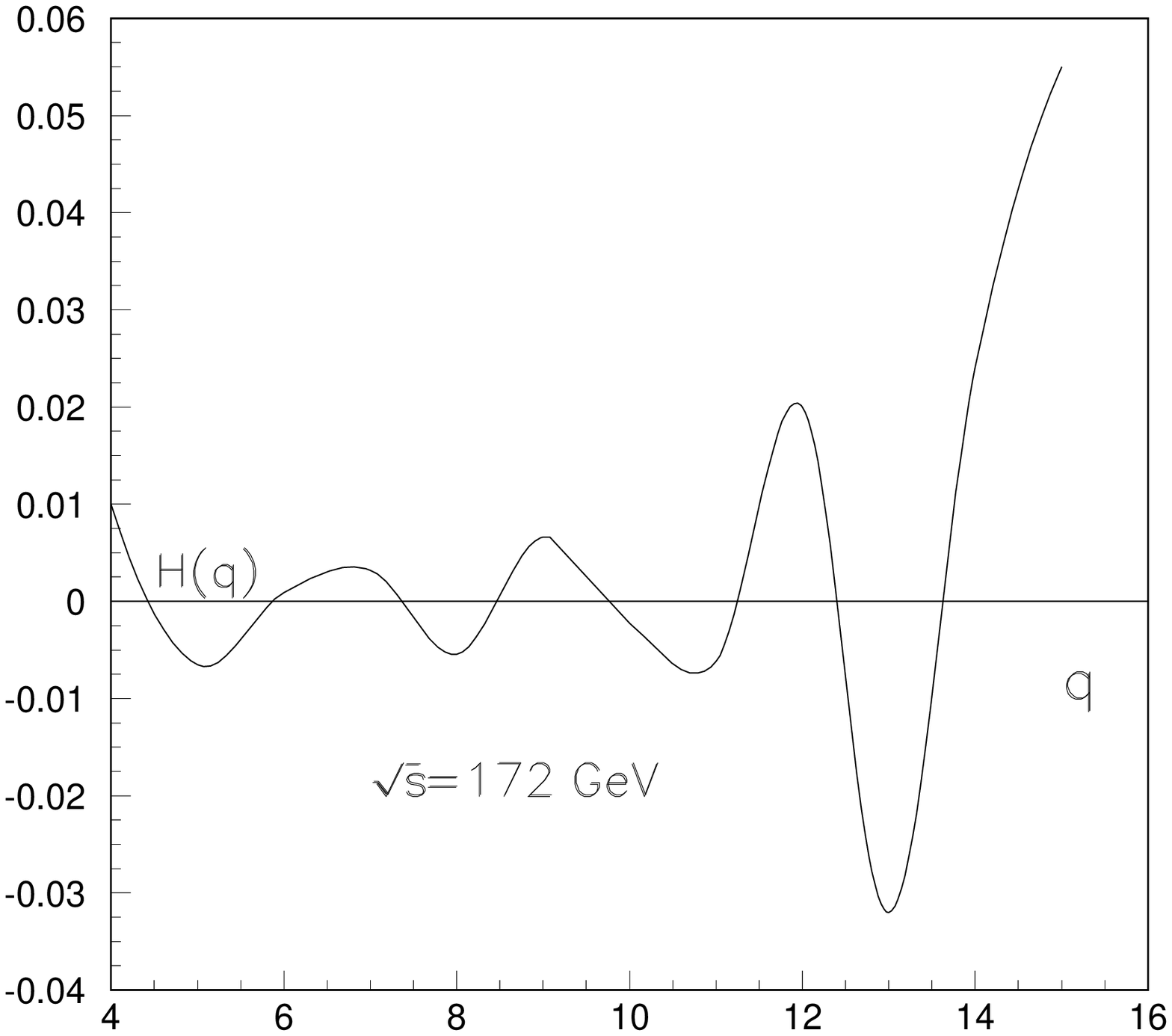}
\caption{$H_q$ at 172GeV.}
\label{14kdfig}
\end{minipage}\hfill
\begin{minipage}[b]{.3\linewidth}
\centering
\includegraphics[width=\linewidth, height=2in, angle=0]{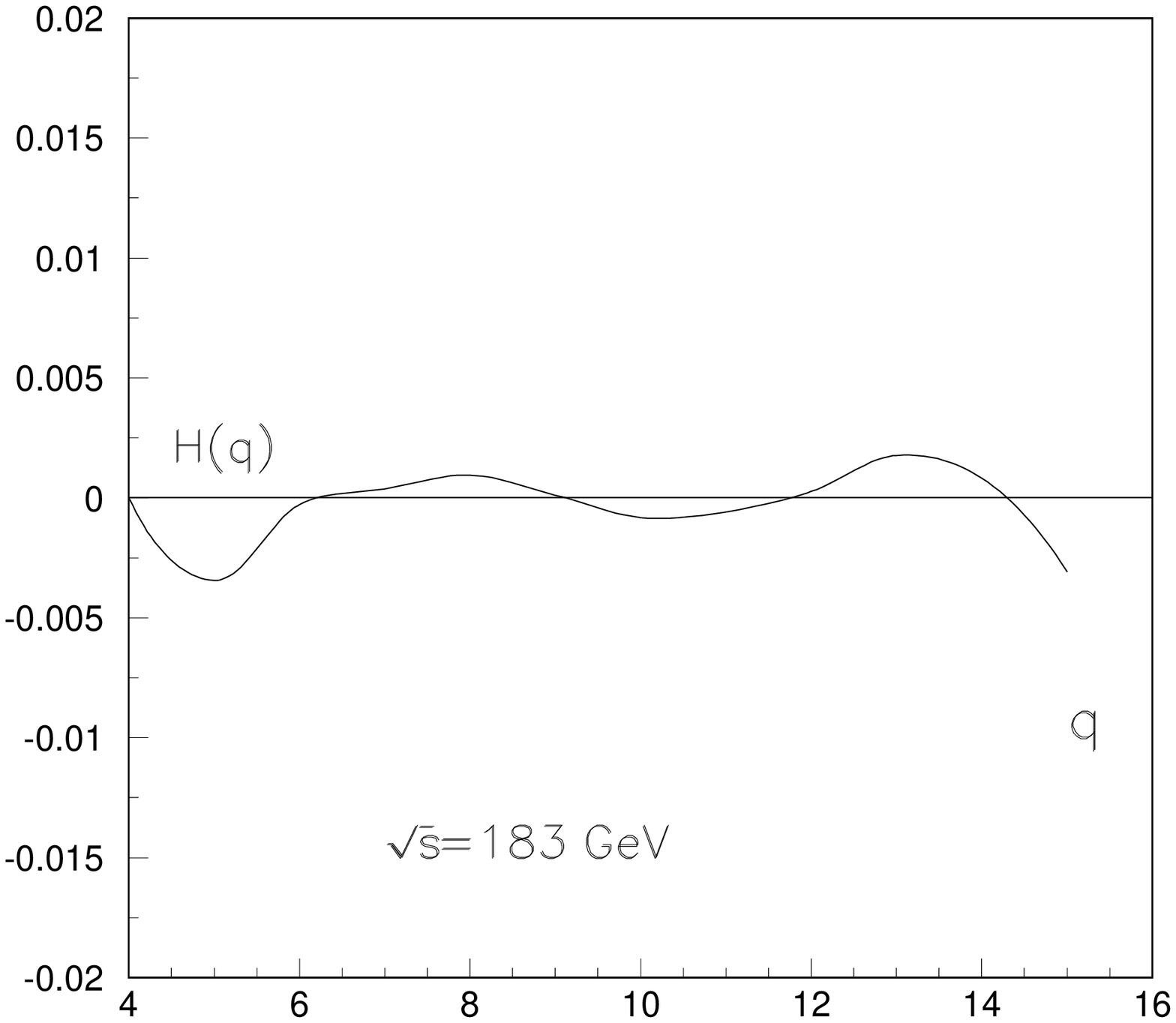}
\caption{$H_q$ at 183GeV.}
\label{15kdfig}
\end{minipage}\hfill
\begin{minipage}[b]{.3\linewidth}
\centering
\includegraphics[width=\linewidth, height=2in, angle=0]{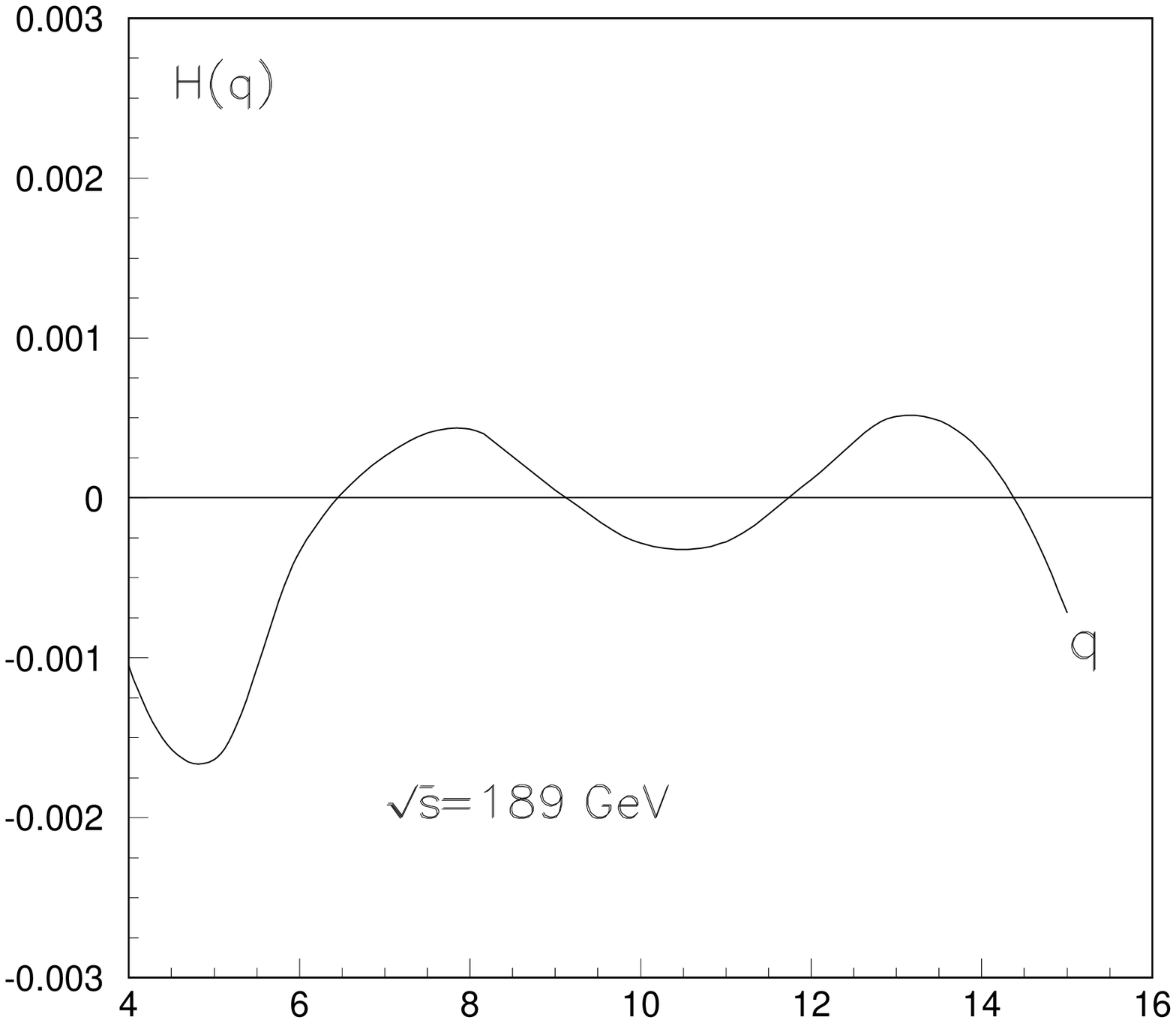}
\caption{$H_q$ at 189GeV.}
\label{16kdfig}
\end{minipage}
\end{figure}
\end{document}